\documentclass[iop]{emulateapj}

\usepackage{natbib}
\newcommand       \kpc		{\,{\rm kpc }}
\newcommand       \kms		{\,{\rm km\,s^{-1} }}
\newcommand       \erg		{\,{\rm erg }}
\newcommand       \s		{\,{\rm s }}
\newcommand       \pc		{\,{\rm pc }}
\newcommand       \hii          {\ion{H}{2} }
\newcommand       \micronm       {\,\mu{\rm m} }

\shorttitle{MASSIVE STAR FORMING REGIONS IN THE GALAXY}
\shortauthors{Rahman \& Murray}

\begin{document}
\title{Massive Star Forming Regions in the Galaxy using the Spitzer
  GLIMPSE Survey} \author{Mubdi Rahman\altaffilmark{1} and Norman
  Murray\altaffilmark{2,3}} \altaffiltext{1}{Department of Astronomy
  \& Astrophysics, University of Toronto, 50 St. George Street,
  Toronto, Ontario, M5S 3H4, Canada} \email{rahman@astro.utoronto.ca}
\altaffiltext{2}{Canadian Institute for Theoretical Astrophysics,
  University of Toronto, 60 St. George Street, Toronto, Ontario, M5S
  3H8, Canada}
\altaffiltext{3}{Canada Research Chair in Astrophysics}
 
\begin{abstract}
  We examine the thirteen most luminous sources in the WMAP free-free
  map using the Spitzer GLIMPSE and MSX surveys to identify massive
  star formation regions, emitting one-third of the Galactic free-free
  luminosity. We identify star forming regions by a combination of
  bubble morphology in 8 $\micronm$ (PAH) emission and radio
  recombination line radial velocities. We find 40 star forming
  regions associated with our WMAP sources, and determine unique
  distances to 31. We interpret the bubbles as evidence for radial
  expansion. The radial velocity distribution for each source allows
  us to measure the intrinsic speed of a region's expansion. This
  speed is consistent with the size and age of the bubbles. The high
  free-free luminosities, combined with negligible synchrotron
  emission, demonstrate that the bubbles are not driven by supernovae.
  The kinetic energy of the largest bubbles is a substantial fraction
  of that measured in the older superbubbles found by Heiles. We find
  that the energy injected into the ISM by our bubbles is similar to
  that required to maintain the turbulent motion in the gas disk
  inside 8 kpc. We report a number of new star forming regions powered
  by massive ($\textrm{M}_{*} > 10^4 \textrm{M}_\sun$) star clusters.
  We measure the scale height of the Galactic O stars to be
  $h_{\textrm{*}} = 35 \pm 5 \pc$. We determine an empirical
  relationship between the PAH and free-free emission of the form
  $F_{\textrm{PAH}} \propto F^2_{\textrm{ff}}$. Finally, we find that
  the bubble geometry is more consistent with a spherical shell rather
  than a flattened disk.
\end{abstract}

\keywords{infrared: ISM -- ISM: \hii regions -- stars: formation}

\section{Introduction}

Massive star forming regions provide a unique laboratory to study both
the process of massive star formation and feedback within the
Galaxy. These clusters are home to the most massive stars in the
Galaxy, including the vast majority of O and B stars. The stars in
turn produce most of the ionizing luminosity and stellar winds that
inject energy and momentum into the interstellar medium (ISM), blowing
bubbles and producing shell structures. Finally, the most massive
stars explode as supernovae, which also inject energy and momentum
into the ISM.

The most massive star clusters (M$_{*}> 10^4 \textrm{ M}_{\sun}$),
referred to as {\it super star clusters}, have been regularly observed
in extragalactic star forming regions \citep{ho97}, but until recently
such massive clusters have evaded detection in our own Galaxy. This
has largely been due to the heavy dust obscuration within our own
Galactic disk. Over the last decade a number of young, massive
clusters have been found in the Galaxy, including the Arches and
Quintuplet clusters near the Galactic centre region \citep{figer99},
Westerlund 1 \citep{clark05}, and RSGC 1, 2, \& 3 \citep{figer06,
  davies07, clark09}, either by directly imaging the stars, or by the
stellar radio emission.

Another possible method of locating these young, massive star forming
regions is to look for the environmental effects caused by such
clusters, such as \hii regions, or shell and bubble
structures. These effects can be observed in wavebands where extinction
through the Galactic plane becomes less of an issue, such as the radio
and infrared. Extensive surveys of \hii regions both in the northern
and southern sky have been conducted with limited success in finding
these massive star forming regions \citep[for a recent census, refer
to][]{conti04}.

In \citet[][hereafter Paper I]{paperI}, we used the Wilkinson
Microwave Anisotropy Probe (WMAP) maximum entropy method free-free
foreground emission map \citep{bennett03, gold09} to determine the
star formation rate in the Galaxy. We found that the 18 most luminous
regions (located within 13 WMAP sources) produce over half the total
ionizing luminosity of the Galaxy, implying that half the total number
of O stars reside in these regions. These sources contain bubble
structures having radii ranging from 5 to 100 pc. Most of the ionizing
photons from the embedded stellar population escape the bubble and
ionize the surrounding material, creating the Extended Low-Density
regions (ELD) \citep{lockman76, anan85, anan85a}. Examination of the
Spitzer Galactic Legacy Infrared Midplane Survey Extraordinaire
(GLIMPSE) 8 $\mu$m images shows that many known \hii regions appear on
shells or walls of bubbles. These known \hii regions are generated
either by the illumination of swept up material by the central
cluster, or by star formation triggered in the swept up material. In
either case, the luminosity of the central cluster is larger than that
given by summing the total emission from all of the \hii regions in
the area.

Further, a large number of the known \hii regions in a WMAP source had
not previously been associated with one another, a result of their
disparet radial velocities. The range of radial velocities had been
translated into different distances along the line of sight. We, on
the other hand, interpret the differences in radial velocity as the
result of bubble expansion; the \hii regions lie on shells, with
velocity differences of order $15\kms$, consistent with the expected
expansion speed of a bubble in the ISM, e.g., \citet{harperclark09}.

In this paper we analyze the most luminous WMAP sources using the
GLIMPSE and MSX surveys, and previously known \hii region
velocities. We describe our procedure for associating the \hii and PAH
emission using an intrinsic expansion velocity criteria in \S
\ref{dataanalysis}. We discuss the general properties of the star
forming regions (which we define in \S \ref{sec:SFR}) in \S
\ref{sfr}. We discuss each of the star forming regions in depth in
Section \ref{individual}. In Section \ref{discussion}, we determine
the scale height of O stars in the Galactic disk, quantify the
relationship between the PAH and free-free emission, discuss the
expansion of the star forming regions as a turbulent driving mechanism
of the Galaxy's molecular gas, and comment on the three-dimensional
geometry of observed bubble structures. Finally, we summarize our
results in Section \ref{summary}.

\section{Data Analysis }
\label{dataanalysis}
In Paper I, we divided the total flux of the WMAP regions along a
given line of sight based on various distance determinations (radio
recombination lines, molecular absorption lvines, and stellar
distances).  Using this division, one-third of the Galactic free-free
emission arises from 14 discrete WMAP sources.  These sources contain
more than one-third of the O star population of the Galaxy.

Here we investigate in more detail, using additional radial
velocities, the 13 most luminous free-free sources in the Galaxy from
Paper I (we exclude the Galactic Centre region since it is so well
studied already).

\begin{figure}
  \includegraphics[width=7in]{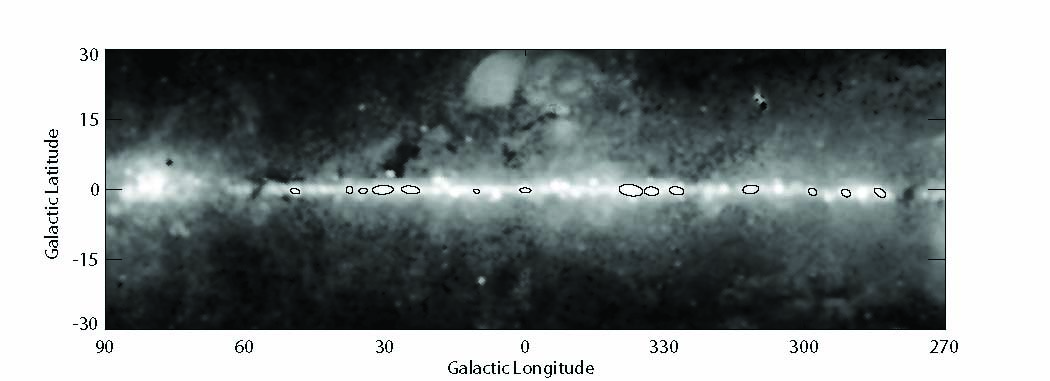}
  \caption{ The WMAP free-free foreground emission map with the target
    sources indicated with black ellipses. The map is described in
    Paper I.}
\label{wmapimage}
\end{figure}

Each of the WMAP sources are fit by ellipses, with major axes between
$\sim$2 to 5 degrees; sizes, positions and fluxes are given in Table
\ref{wmap}. These sources are confused (contain multiple star forming
regions) as a result of the large beam size of the WMAP satellite. The
WMAP free-free map is presented in Figure \ref{wmapimage}, indicating
the location of each of the thirteen sources.

We use published radial velocity measurements together with GLIMPSE 8
$\mu$m images to better resolve and classify these star forming
regions. This improves on our use of the catalogue of
\citet{russeil03} in Paper I. In other words, we classify star forming
regions on the basis of kinematic distances (as determined through
radial velocities towards known \hii regions) and morphology seen in
the 8 micron PAH bands.

We carried out a SIMBAD search for \hii regions within each of the
WMAP sources and compiled the associated hydrogen recombination line
velocities.

Our morphological analysis was primarily conducted using the 3.1 by
2.4 degree Band 4 mosaic images from GLIMPSE \citep{benjamin03}.
In many cases, these mosaics were insufficiently large to encompass
the entire WMAP sources from Paper I.  In these cases, adjacent images
were mosaicked together using the Montage package. In cases where the
sources were outside the GLIMPSE coverage, we substituted Band A
mosaics from MSX \citep{price01}.

We use the 8 micron band because it is dominated by PAH emission,
which \citet{cohen01} have shown to trace the free-free emission
reasonably well. We discuss and expand upon this relationship in depth
in \S \ref{pah}.

\subsection{Definition of Star Forming Regions }
\label{sec:SFR}
We define star forming regions (SFRs) as subsets of WMAP sources
having similar radial velocities ($\pm 15$ km s$^{-1}$), copious 8
$\mu$m emission in Spitzer and/or MSX images, and finally,
morphologies that are consistent with physical association, e.g.,
bubble, shell, or finger/pillar structures. In other words, star forming
regions are bubbles identified identified by features correlated
both in space and in velocity. 

\begin{figure}
  \includegraphics[width=7in]{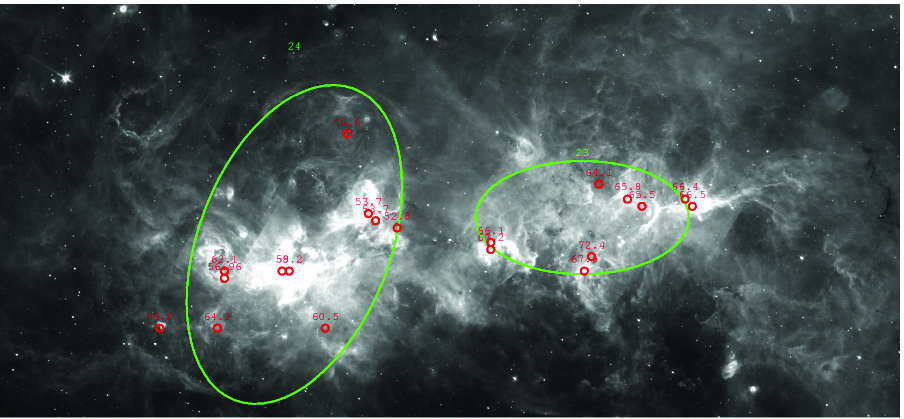}
  \caption{A Spitzer GLIMPSE image of the G49 WMAP source with the two
    star forming regions indicated with the green ellipses and the
    \hii region velocities indicated with the red circles.  (A colour
    version of this figure is available in the online
    journal.) \label{g49vels}}
\end{figure}

To determine the location
and size of each of the star forming regions we overlaid the positions
and velocities of the known \hii regions onto the 8 micron mosaics of
the region and visually inspected the result. An example
classification based on morphology and velocity is shown in Figure
\ref{g49vels}.

Having determined the size and location of a star forming region, we
calculated a median velocity from the measured velocities. A kinematic
distance was then determined using the \citet{clemens85} Galactic
rotation curve. In cases where the kinematic distance is ambiguous, we
used published absorption line data, where available, to determine if
the region is at the ``near'' or ``far'' distance.

In the next section, we describe each of the ``star-forming
regions''. In the supplemental online material, we present tables of
each of the known \hii regions with positions and velocities that are
associated with each of the star forming regions.

\section{Properties of Star Forming Regions }
\label{sfr}
The combined analysis of the WMAP sources with the GLIMPSE image
mosaics and the \hii region velocities yields 40 star forming
regions. Ten of the thirteen WMAP sources harbor multiple star forming
regions; sources within the inner third of the Galaxy are found to be
especially confused, with multiple SFRs.  However, as already noted,
combining morphology from the $8 \mu$m PAH emission with the
velocities of known \hii regions allows for a reasonably clear view of
the physical separations of the star forming regions.

\begin{figure}
  \includegraphics[width=7in]{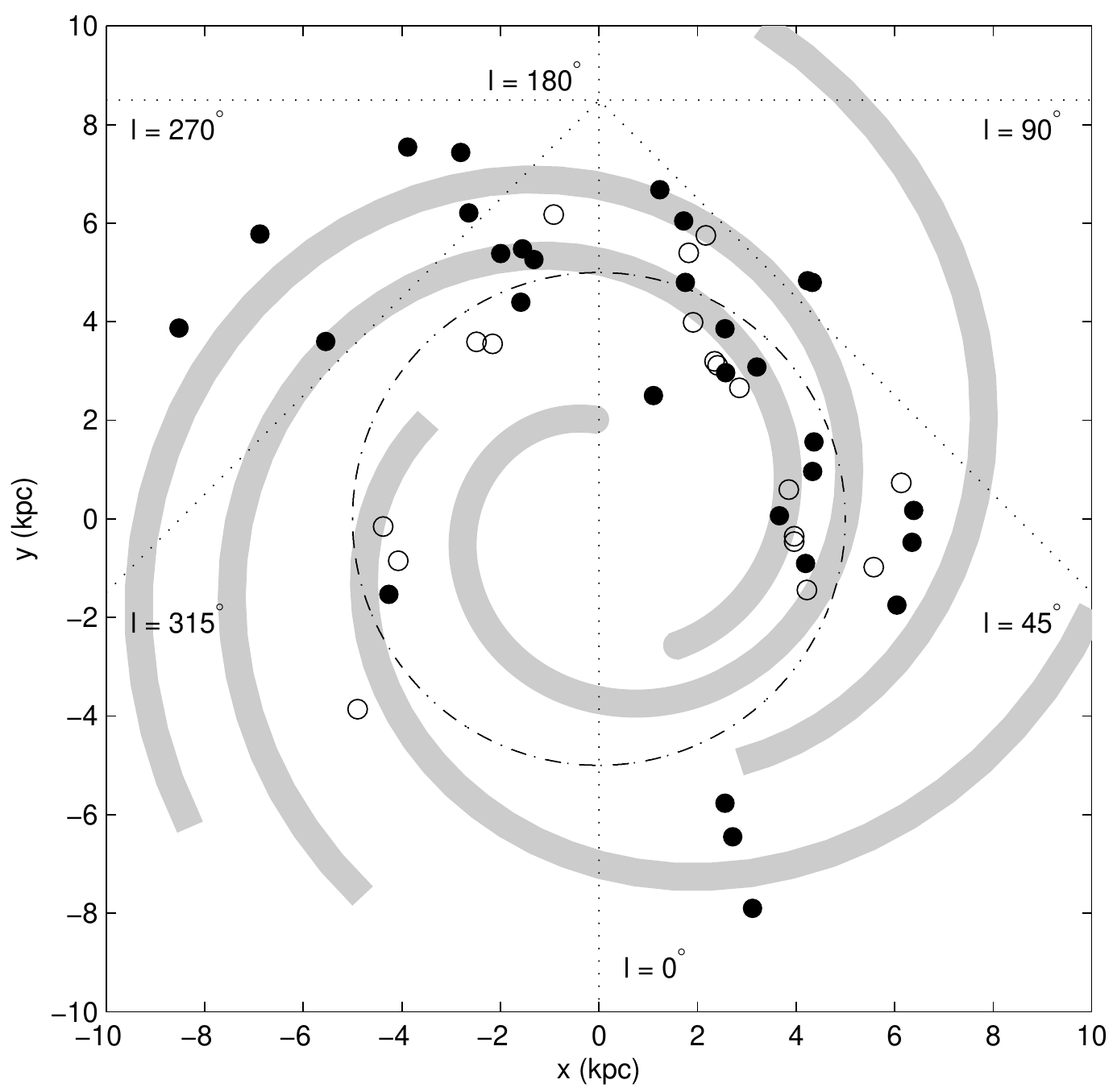}
  \caption{The Galactic distribution of the Star Forming Regions
    identified in this analysis. The filled circles indicate sources
    for which a unique distance has been determined, while open
    circles represent sources for which the kinematic distance is ambiguous
    remains. Both the near and far distances of these sources are presented.
    The spiral arms overlaid are taken from \citet{naka06}.
    \label{regionsmap}}
\end{figure}

These star forming regions have mean radii between an arcminute and a
degree, and distances between 3 and 17 kiloparsecs. The angular sizes
translate into physical radii between 3 to 75 parsecs. The Galactic
distribution of the star forming regions is presented in Figure
\ref{regionsmap}.

The majority of known \hii regions in the star forming regions lie in
shell-like structures traced by PAH emission in the Spitzer or MSX
images.  Most of the remainder are found in the centre of the SFR.

We interpret those \hii regions lying on shell-like structures in one
of two ways. In many cases it is clear that the \hii region is simply
swept up shell material illuminated by a central cluster. In other
cases, however, the shell material is illuminated by young stars
resulting from triggered star formation in the swept up shell, as
described by \citet{elmegreen98}. In some cases, the triggered stars
are blowing their own bubbles in the shell.

In many cases, we find a low level of emission towards the centre of
the bubble or shell structure; we interpret this as the evacuation of
the material from the interior of the bubble.

Although a comprehensive search for known \hii regions was conducted
over each of the WMAP sources, not every emission feature in the 8
micron GLIMPSE image is associated with a radial velocity
determination. In some cases, particularly bright PAH features in the
GLIMPSE mosaics are not coincident with any radial velocity
measurements, indicating that we are potentially missing star forming
regions within the WMAP sources. We identify these regions in the
description of each of the WMAP sources, and we suggest that further
radial velocity measurements would be instructive. This does add an
uncertainty in determining the true stellar population of such regions
and we comment where this is the case. In some cases we believe that a
significant star forming region has not been identified due to a lack
of conclusive \hii data.
  
We present our SFRs in Table \ref{sfrlist} with the columns as
follows: column (1) the catalogue number, columns (2) and (3) the
position of the sources in Galactic coordinates, columns (4) and (5)
the semimajor and semiminor axes of the regions in arcseconds, column
(6) the position angle of the region, column (7) the median velocity
in km s$^{-1}$, columns (8) and (9) the measured and corrected
half-spread of velocities in km s$^{-1}$ where sufficient \hii regions
are identified as associated with the star forming region, column (10)
the distance to the region in kiloparsecs, column (11) the mean radius
of the source in parsecs, column (12) the dynamical age of the region
given the velocity spread, and column (13) the reference that resolves
the distance ambiguity (where applicable). In cases where the
kinematic distance ambiguity remains, we indicate the properties
resulting from both distances in columns (10), (11), and (12). We
refer to regions from this table as ``SFR'' followed by the catalogue
number hereafter.

\begin{figure}
  \includegraphics[width=7in]{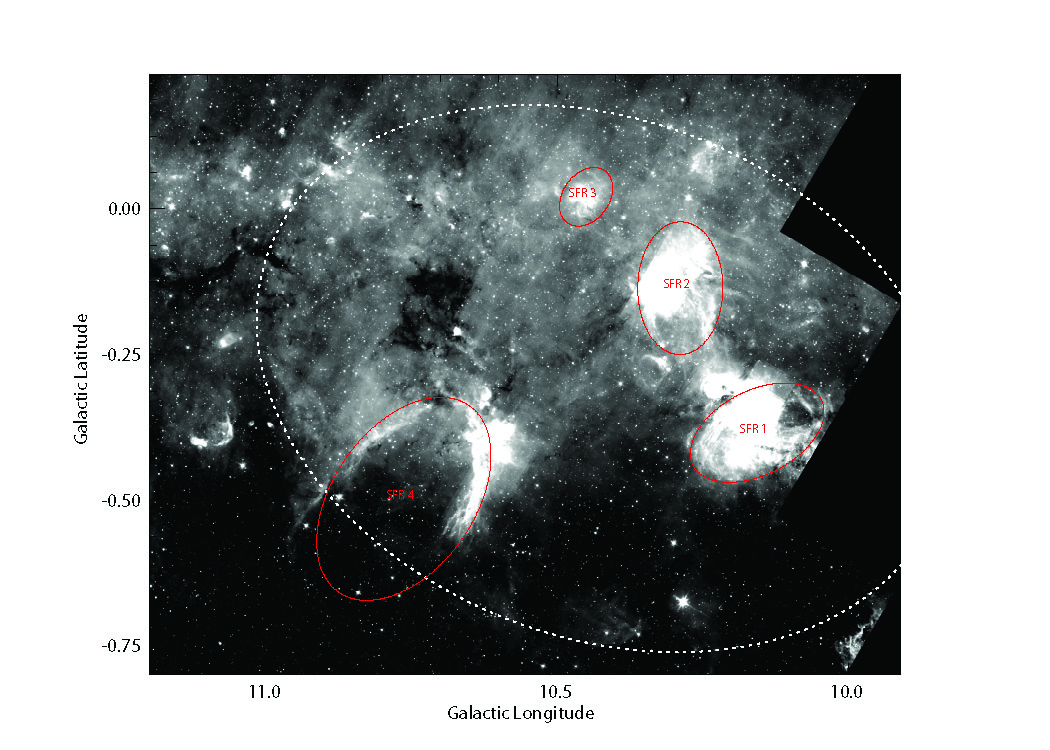}
  \caption{The 8$\mu$m GLIMPSE Image of the star forming regions in
    G10 \label{G10glimpse}}
\end{figure}

\begin{figure}
  \includegraphics[width=7in]{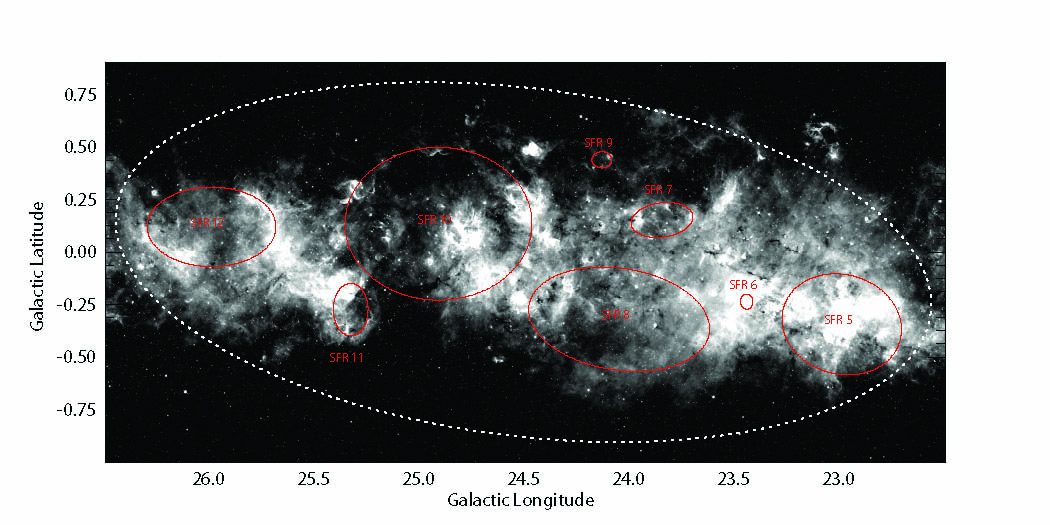}
  \caption{The 8$\mu$m GLIMPSE Image of the star forming regions in
    G24 \label{G24glimpse}}
\end{figure}

\begin{figure}
  \includegraphics[width=7in]{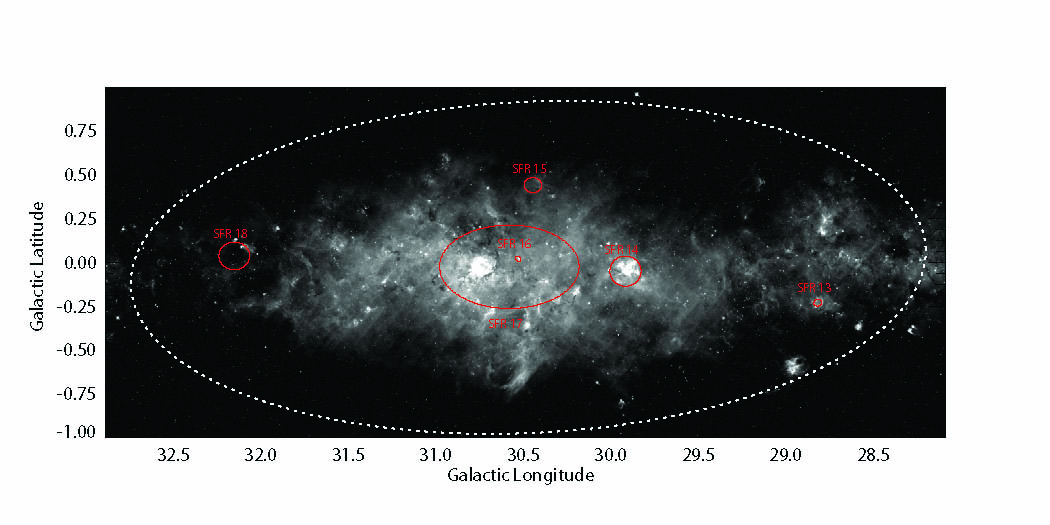}
  \caption{The 8$\mu$m GLIMPSE Image of the star forming regions in
    G30 \label{G30glimpse}}
\end{figure}

\begin{figure}
  \includegraphics[width=7in]{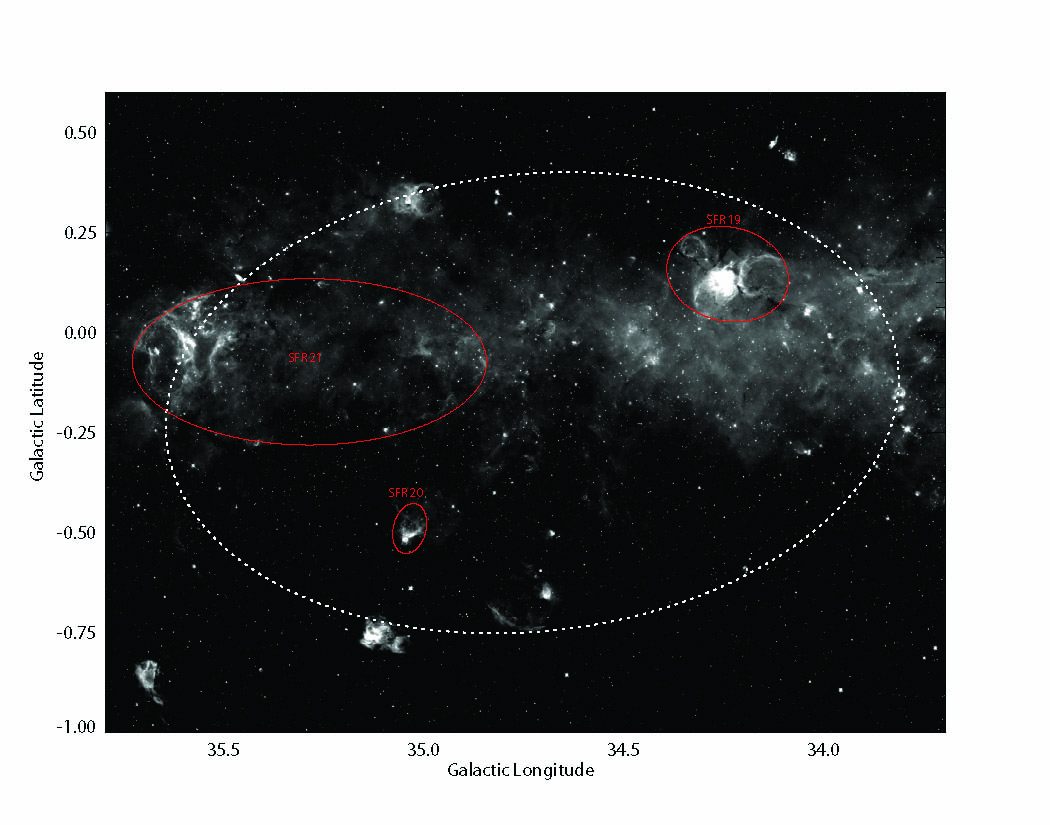}
  \caption{The 8$\mu$m GLIMPSE Image of the star forming regions in
    G34 \label{G34glimpse}}
\end{figure}

\begin{figure}
  \includegraphics[width=7in]{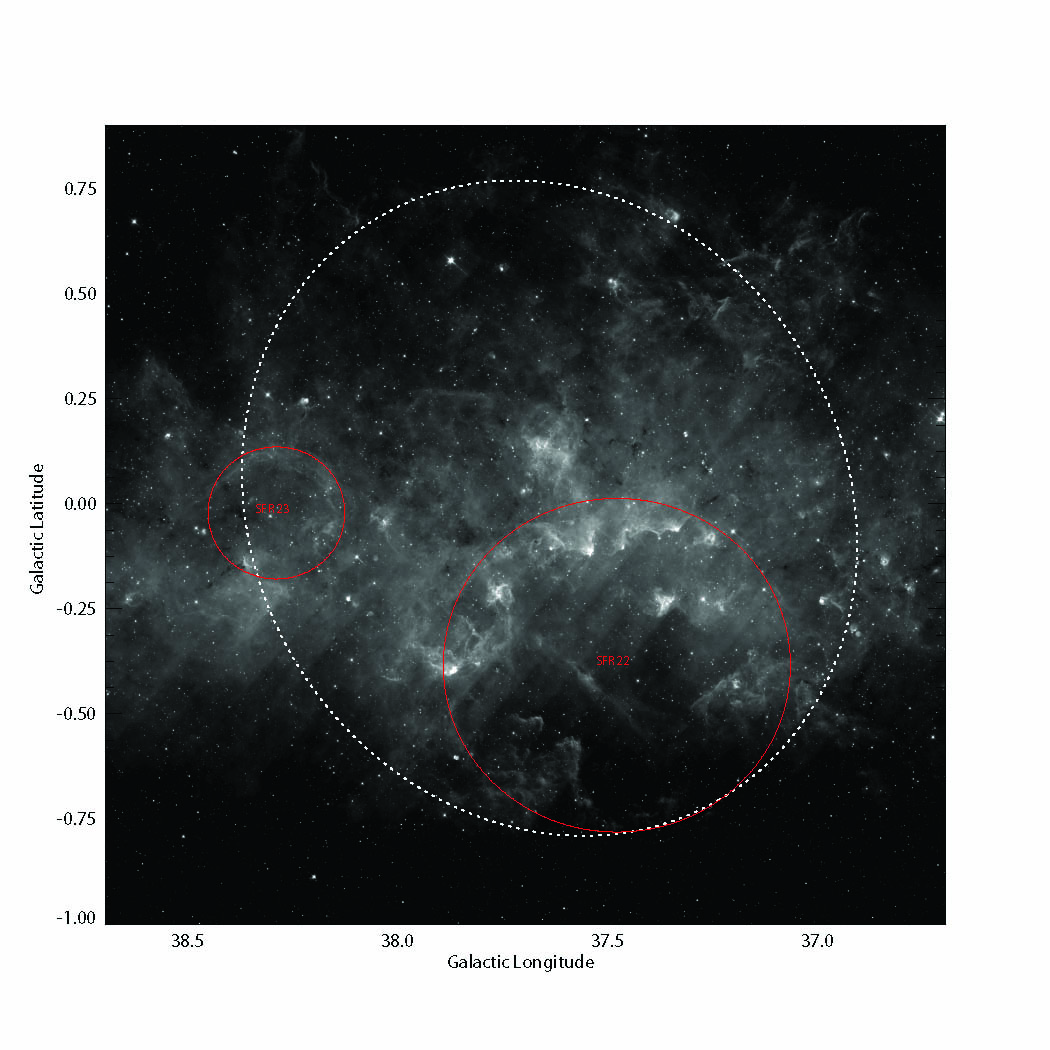}
  \caption{The 8$\mu$m GLIMPSE Image of the star forming regions in
    G37 \label{G37glimpse}}
\end{figure}

\begin{figure}
  \includegraphics[width=7in]{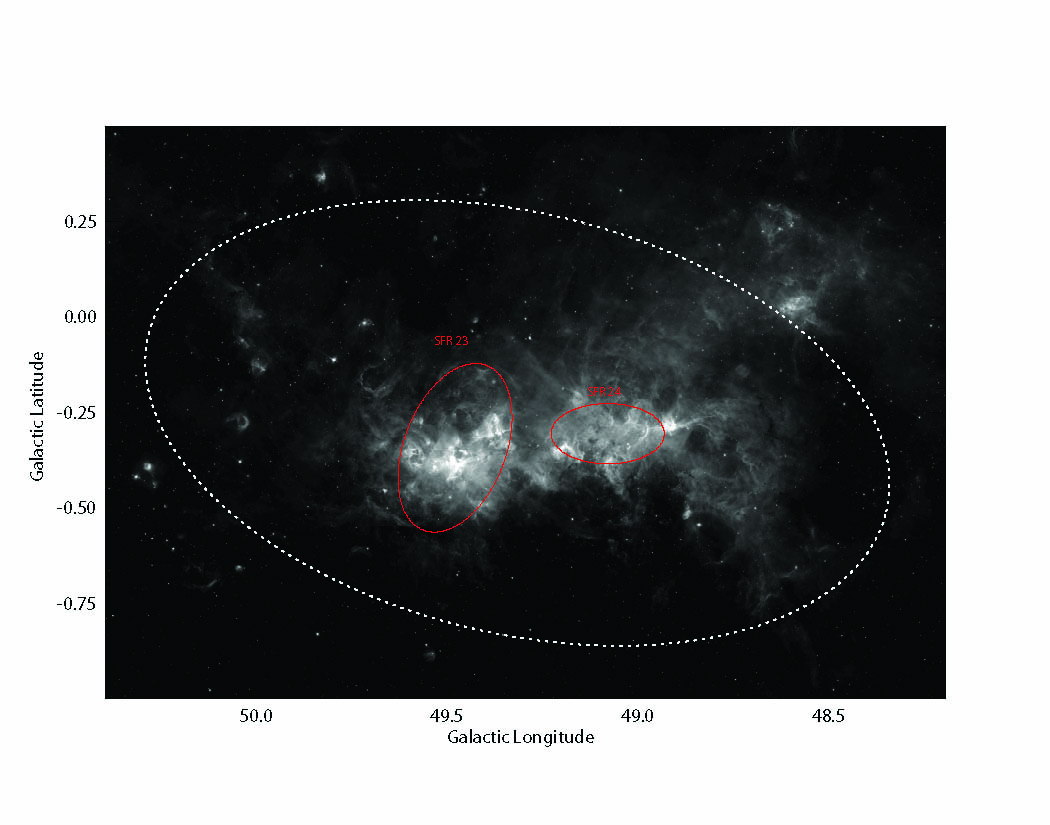}
  \caption{The 8$\mu$m GLIMPSE Image of the star forming regions in
    G49 \label{G49glimpse}}
\end{figure}

\begin{figure}
  \includegraphics[width=7in]{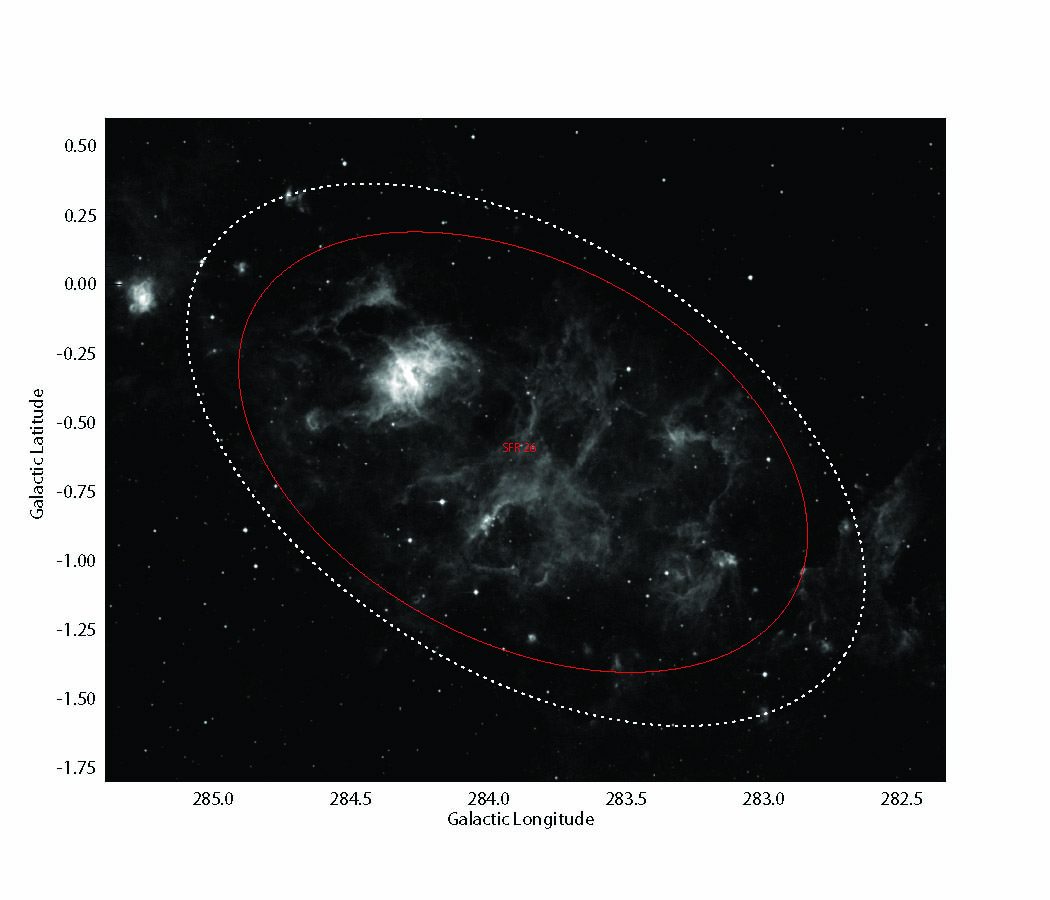}
  \caption{The 8$\mu$m MSX Image of the star forming regions in G283
    \label{G283msx}}
\end{figure}

\begin{figure}
  \includegraphics[width=7in]{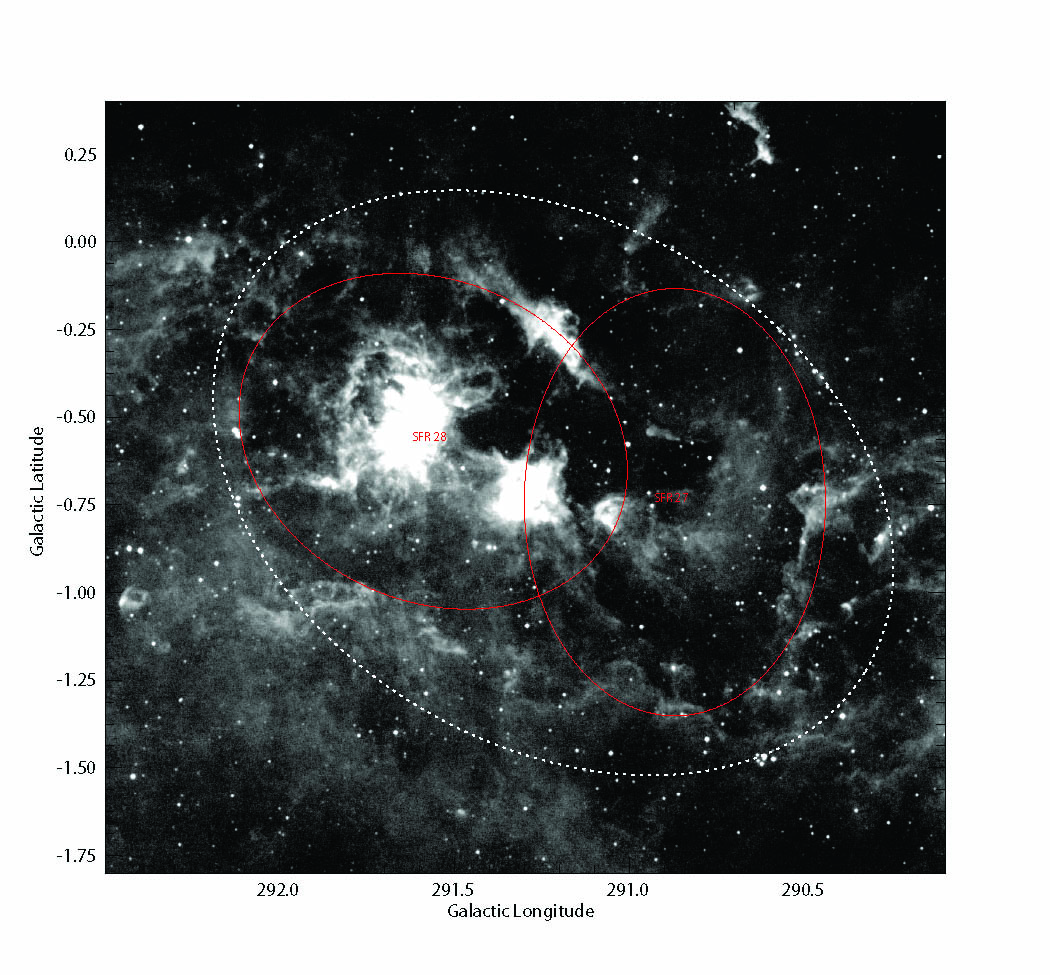}
  \caption{The 8$\mu$m MSX Image of the star forming regions in G291
    \label{G291msx}}
\end{figure}

\begin{figure}
  \includegraphics[width=7in]{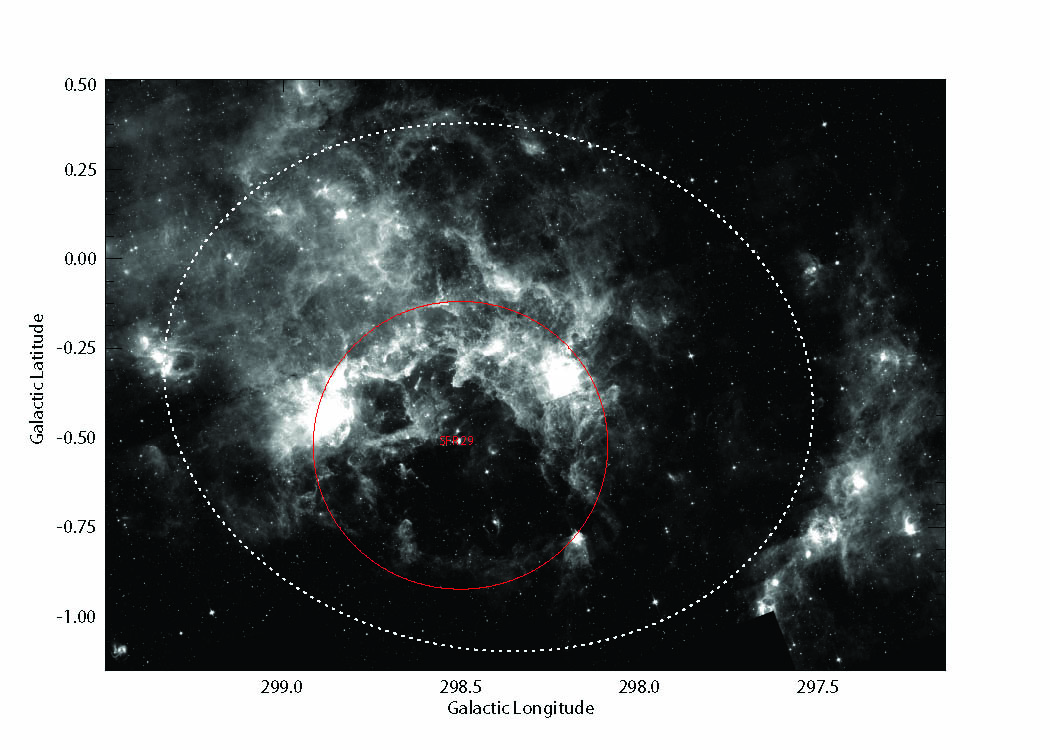}
  \caption{The 8$\mu$m GLIMPSE Image of the star forming regions in
    G298 \label{G298glimpse}}
\end{figure}

\begin{figure}
  \includegraphics[width=7in]{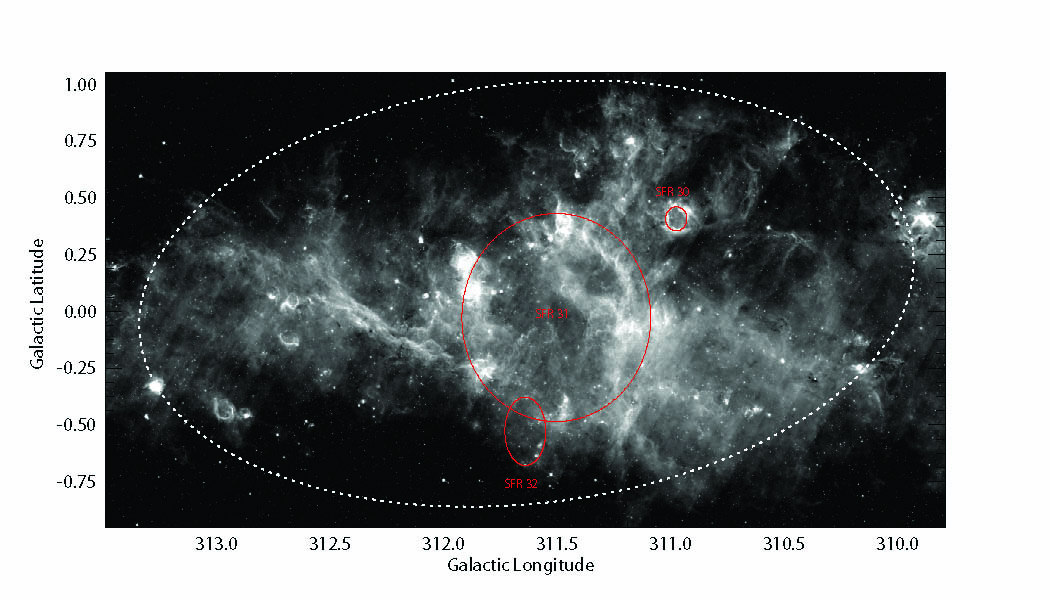}
  \caption{The 8$\mu$m GLIMPSE Image of the star forming regions in
    G311 \label{G311glimpse}}
\end{figure}

\begin{figure}
  \includegraphics[width=7in]{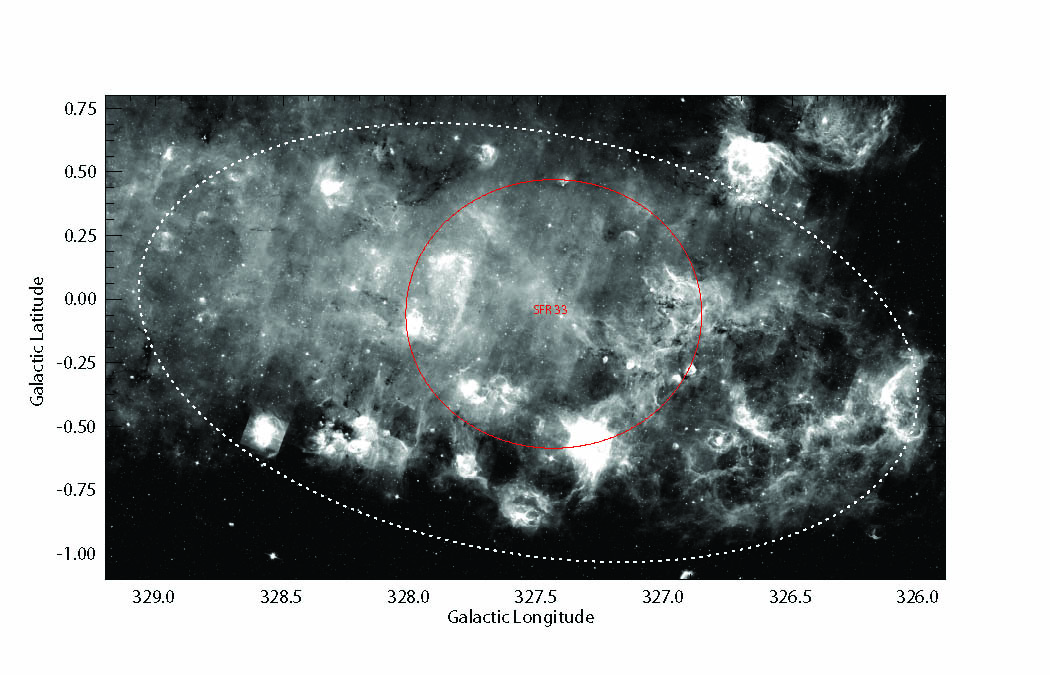}
  \caption{The 8$\mu$m GLIMPSE Image of the star forming regions in
    G327 \label{G327glimpse}}
\end{figure}

\begin{figure}
  \includegraphics[width=7in]{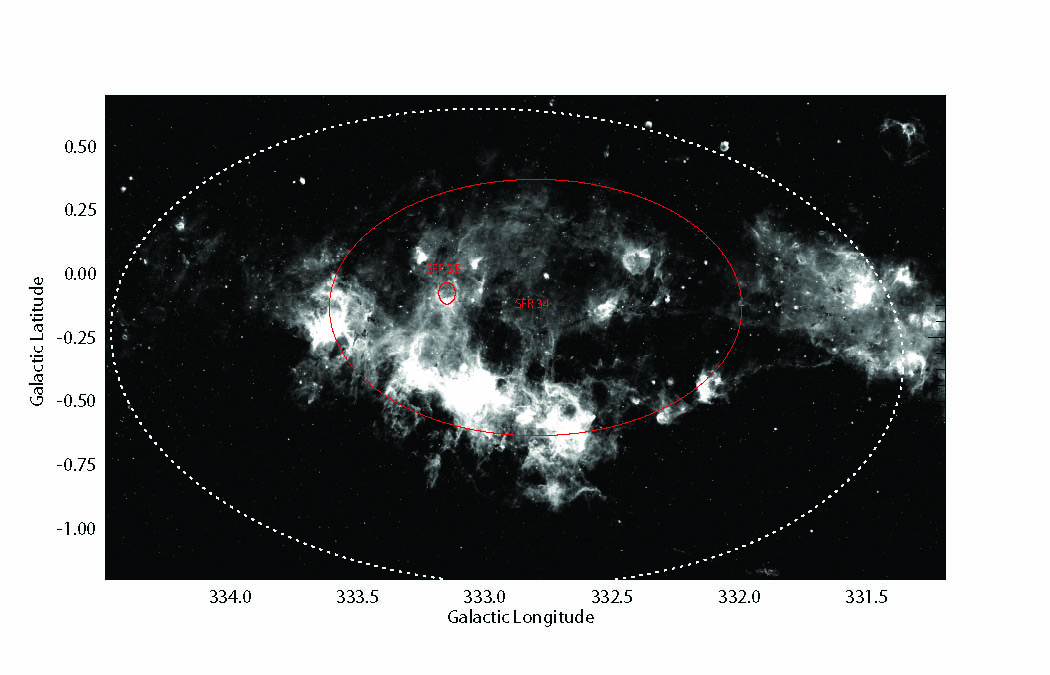}
  \caption{The 8$\mu$m GLIMPSE Image of the star forming regions in
    G332 \label{G332glimpse}}
\end{figure}

\begin{figure}
  \includegraphics[width=7in]{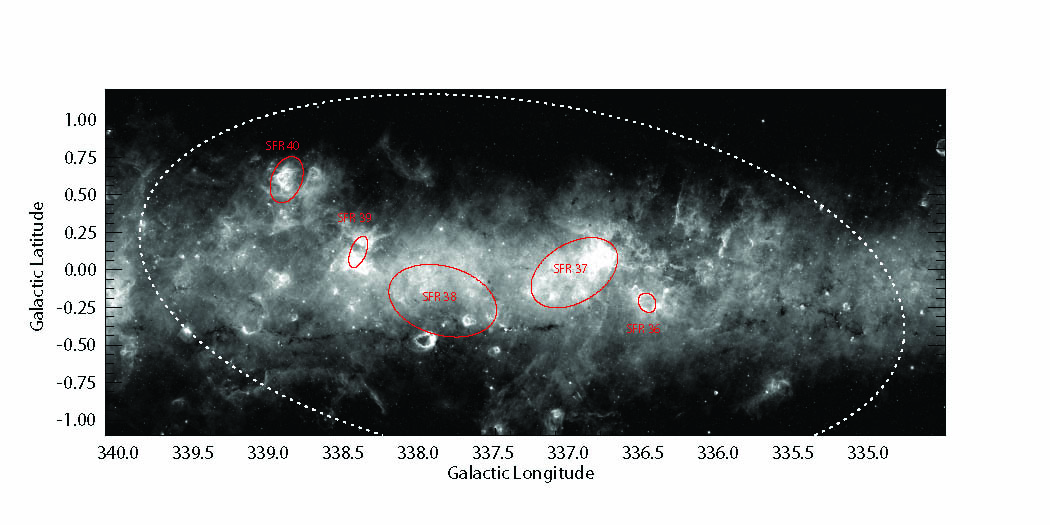}
  \caption{The 8$\mu$m GLIMPSE Image of the star forming regions in
    G337 \label{G337glimpse}}
\end{figure}

We present 8 micron images from GLIMPSE or MSX for each of the WMAP
sources in Figures \ref{G10glimpse} to \ref{G337glimpse}. On each of
these images, the location and size of the WMAP source is indicated
with a dotted ellipse, and the location of SFRs is indicated by a
solid (red, in the online copy) ellipse. Each of these images is
presented with a logarithmic stretch to highlight fainter
morphological features.

While individual velocities towards \hii regions may be known with
great precision (generally to better than $\pm$ 1 km s$^{-1}$),
velocities to the entire regions are determined with much less
accuracy; the \hii regions are randomly distributed (in angle) around
sources. Further, some of the SFRs are associated with a very limited
number of radial velocities measurements. In either case, the maximum
error in the mean source velocity is constrained by our selection
criteria that a region have a velocity spread no greater than
$\pm15\kms$. This maximal velocity error corresponds to an error in
the kinematic distances of approximately 30\%.

To estimate the half-velocity spread, $\Delta v_{\rm m}$, we first
require that a SFR have multiple velocity measurements
available. Next, we require that there be a spread in the measured
velocities greater than $3\kms$; otherwise we suspect that what appear
to be multiple velocity measurements are actually measurements of the
same \hii region (the Galactic coordinates are often not sufficiently
precise to discriminate between physically different versus apparently
different sources). Given a region that satisfies these two criteria,
we define the {\em measured} $\Delta v_{\rm m}$ as half the difference
between the maximum and minimum radial velocity.

This parameter must be corrected for sampling effects; it is unlikely
that the maximum and minimum velocity of an expanding region will be
found in a random selection of \hii regions.  Further, since most of
the known \hii regions appear on the edges of (presumably expanding)
shells, the component of the expansion velocity projected onto the
radial velocity is expected to be smaller than unity. Thus the
measured half-velocity spread would generally be an underestimate of
the actual expansion velocities of the regions.

We correct the half-velocity spread by a geometric factor found by
running a Monte Carlo simulation modeling the difference between the
observed half-velocity spread and the actual expansion velocity with a
given number of velocity observations. The mean correction factor is
2.0, 1.4 and 1.2 for 5, 10 and 15 velocity measurements
respectively. The error on this factor ranges from a factor of 2 for 5
velocity measurements, to 15\% for 15 measurements. The {\em
  corrected} half-velocity spreads, $\Delta v_{\rm c}$, are also given
in Table \ref{sfrlist}. For regions where less than 5 velocity
measurements are available, no correction factor is applied as the
Monte Carlo simulation shows that the result is unreliable.  These
regions do not have a corrected half-velocity spread presented. The
measured half-velocity spreads for these regions should be treated as
lower limits of the expansion velocity.

The dynamical age, $\tau_{dyn}$, of the system is determined by
dividing the mean radius of the regions by the half-velocity spread.
The dynamical ages of all of the regions are below 10$^7$ years,
consistent with the existence of massive main-sequence stars producing
large amounts of ionizing luminosity.

The true age of each of these systems is an undetermined quantity but
the selection criteria for these regions imposes some constraint on
the ages. Specifically, these regions are selected as the most
luminous in free-free emission, indicating the presence of massive O
stars, specifically those with short main-sequence lifespans. In Paper
I we showed that the total ionizing luminosity of a cluster with a
standard initial stellar mass function rapidly decreases after the
cluster reaches an age of 4 Myrs. This indicates that the majority of
the ionizing luminosity in a young cluster is quenched even before the
late-O stars evolve off the main sequence.

In SFRs where a discernible, closed bubble is illuminated with
free-free or PAH emission, an upper limit to the true age exists.  In
SFRs with only a diffuse bubble structure there is an ambiguity; the
bulk of the ionizing luminosity may come from a central cluster, or it
may come from star clusters that have been triggered along the shell
of the region evacuated by a central cluster. In the latter case, it
would be incorrect to constrain the age of the region to be less than
$4$ Myrs.  This circumstance is particularly relevant when searching
for a central stellar cluster; if the age cannot be constrained to
under 4 Myr, the most massive stars in the central cluster have
already evolved off the main sequence, leaving more evolved objects
rather than bright, blue main-sequence objects.

\subsection{Bubbles}

In many of these regions, we observe bubbles at a variety of
scales. Some bubbles essentially enclose the associated star forming
regions, while smaller bubbles often lie in the walls of larger
bubbles. Many of the smaller bubbles are listed in
\citet{churchwell06}, but the larger bubbles, with radii larger than
10', have not been previously catalogued.

In Paper I, we find that each of the WMAP sources that we have
selected require ionizing luminosities of Q$_{0} > 10^{51}$ s$^{-1}$,
consistent with young stellar populations with M $> 10^{4}$
M$_{\sun}$. Such massive clusters may drive superbubbles, a notion
supported by the fact that the largest bubbles we find have mean radii
approaching $50\pc$.

We present a list of bubbles associated with each of the star forming
regions in Table \ref{bubblelist}. The columns are as follows: column
(1) the catalogue number, columns (2) and (3) the positions of the
bubbles in Galactic coordinates, columns (4) and (5) the semimajor and
semiminor axes of the bubbles in arcminutes, column (6) the position
angle of the bubble, column (7) the associated star forming region
from Table \ref{sfrlist}, column (8) the mean radius of the bubble in
parsecs using the distances to the associated star forming regions,
column (9) the morphology classification flag of the bubble, and
column (10) the GLIMPSE bubble catalogue number from
\citet{churchwell06}.  Where the kinematic distance ambiguity has not
been resolved, the mean radius of both the near and far distances are
indicated. The morphology classification flag is C if a nearly closed
or complete shell is visible in the GLIMPSE 8 micron image, or B if
the bubble appears to be broken or blown-out. The GLIMPSE bubble
catalogue number is indicated where the bubble has a similar position,
size and shape to the source indicated in \citet{churchwell06}.

\subsection{\hii Region Radial Velocities}

Finally, we list the individual \hii region velocities for each of the
identified star forming regions in Table \ref{hiilist}, available as
an online supplemental table. The columns are as follows: column (1)
the associated star forming region from Table \ref{sfrlist}, column
(2) the name of the \hii region, columns (3) and (4) the positions of
the region in Galactic coordinates, column (5) the velocity of the
region with respect to LSR in km s$^{-1}$, and column (6) the
reference from which the velocity was taken.

\section{Individual WMAP Sources }
\label{individual}
\subsection{G10 including the W31 region }
\label{G10}

This WMAP source is dominated by four seemingly disparate star forming
regions (see Figure \ref{G10glimpse}), although SFR 1 and 2 may be
associated as they are both in W31 \citep{westerhout58}. We begin with
a discussion of these two regions.

SFR 1 and 2 contain the brightest PAH emission, although there is a
significant 8 micron background over this entire WMAP source that we
cannot clearly attribute to any of the 4 regions. In addition to being
in W31, SFR 1 and 2 are also linked by their close angular proximity
and similar radial velocities. We treat them as separate regions as
they are each enclosed by their own shell structures. The high surface
brightness 8$\mu$m emission located in the centre of each of SFR 1 and
2 is indicative of a young central cluster that has not had time to
evacuate its environment. This is consistent with the short dynamical
time determined for SFR 1. Strong molecular line emission is found by
\citet{kim02}, who conclude that there are active interactions between
the forming stellar clusters and their parent molecular clouds. They
also assume a closer distance, associated with the expanding 3 kpc arm
of the Galaxy (with a distance of 6 kpc from LSR), but they do note
that there is evidence that the region is at the further distance.
Assuming the near distance, they estimate star cluster masses of
1.22$\times 10^{4}$ M$_{\sun}$ for SFR 1 and 3.11$\times 10^{3}$
M$_{\sun}$ for SFR 2.

The radio nebula G10.0-0.3 and associated soft gamma repeater are
coincident with SFR 1 with a distance of 14.5 kpc, consistent with our
distance \citep{corbel97, vasisht95}. Further, there exists a luminous
blue variable with a total mass (single or binary) exceeding 190
M$_{\sun}$ consistent with the further distance \citep{vk95,
  eikenberry04}. The region also contains a soft-gamma repeater
\citep{kul93}. \citet{corbel04} suggests that W31, specifically SFR 1,
may be separable along the line of sight into a component in the 3 kpc
spiral arm, and a component at 14.5 kpc.  We find this to be
consistent with SFR 1 being a massive young star forming cluster at a
distance of 14.5 kpc, with shell-like structure evident in the
north-west and south-east edges of the region. Adjusting the values of
\citet{kim02} for the further distance, clusters of masses 7.1$\times
10^{4}$ M$_{\sun}$ and 1.8$\times 10^{4}$ M$_{\sun}$ are required for
SFR 1 and 2 respectively.

SFR 3 and 4 are distinctly separated from SFR 1 and 2 as they both
have substantially different radial velocities; the difference in
velocities being $60\kms$ and $15\kms$ respectively. We view SFR 3 as
a smaller, closer star forming region than 1 and 2. SFR 4, however has
the morphology of a blown out bubble. The distance to this region,
16.7 kpc, implies a large bubble size of 46 pc. The strong PAH
illumination on the northwest side of the region is potentially a
triggered star forming region from the expansion of the now blown out
bubble. However, the bubble structure is illuminated along the shell
in all but the south (in the proposed area of blowout), indicating
that a central ionizing source still exists within the bubble that is
illuminating the southeast side; the interior of the bubble appears
evacuated.  We infer that the central cluster is still sufficiently
young to possesses ionizing stars.

At the location $(l, b) = (10.7, -0.16)$, there appears to be a hole
in the background level of PAH emission. There are no known \hii
regions within this hole, suggesting that this is an extinction
effect. If this is an extinction effect at 8 microns, the column
towards this region must be particularly large. We suspect that this
is an infrared dark cloud (IRDC). In 2MASS \citep{2mass}, there does
not appear to be any significant void of background stars, indicating
that this IRDC is sufficiently distant to have no effect on the local
field star population, but is closer than the source of the PAH
background in this region. More detailed study of this object is
required to establish its existence, let alone determine its physical
properties.

\subsection{G24 including the W41 and W42 Regions }
\label{G24}
WMAP source G24 encompasses a particularly complicated region.  It is
one of the largest WMAP sources (spanning nearly 4 degrees in the
Galactic plane) and contains both W41 and W42 \citep{westerhout58}.

SFR 5 encompasses the supernova remnant W41. In the one square degree
region surrounding W41, velocity measurements suggest the \hii regions
are caused by confusion of spatially separated \hii regions
\citep{leahy08}, consistent with the 4 spiral arms along the line of
sight in this direction. We find SFR 6 associated with one of the
overlapping components.  There is a second known supernova remnant,
G22.7 -0.2 \citep{van78}, also within SFR 5.

In addition to SFR 5 and 6, we identify 6 additional star forming
regions within this WMAP source. For a number of these regions,
specifically SFR 7, 9, and 12, the kinematic distance ambiguity
remains. SFR 7 and 8 are easily distinguished by the presence of \hii
regions along their perimeters. There is a lack of strong central PAH
illumination in both of these star forming regions, indicating that
their centres have been evacuated. In addition, the rims of SFR 7 and
8 are not particularly bright in PAH emission, indicating that the
hypothesized shell has dispersed or that the central cluster is no
longer emitting ionizing radiation. Either circumstance implies that
these two regions are older ($> 4.5$ Myr) and that the majority of the
observed PAH and free-free emission is provided by triggered regions
on the perimeter.

SFR 9 contains only a single \hii region velocity in its centre, but
exhibits a clear bubble morphology. SFR 10 is a much larger region,
with a mean radius of 43 pc. While the brightest PAH regions lie
towards the southwest side of the region, the north and eastern edges
are defined by a clear shell structure. There is a \hii region on the
northern shell that has a similar velocity to those of the brighter
regions to the southwest. In addition, there appears to be a central
component to the PAH emission in this region at ($l$,$b$) =
(24.8$^\circ$, 0.1$^\circ$).

SFR 11 is associated with the W42 radio source but also includes an
extended shell structure to the south and west of the original radio
source with consistent velocity measurements along the
shell. \citet{lester85} argue for the near galactic distance on the
basis that there is minimal extinction towards the source in sulphur
forbidden line emission, disregarding the known absorption
velocities. Independent of the central source distance for the W42
region, an associated \hii region at ($l$,$b$) = (25.41$^\circ$,
-0.25$^\circ$) is also determined to be at the near kinematic distance
\citep{kolpak03}, hence we adopt this distance. \citet{blum00} find a
central cluster with a single O5-6.5 star, indicating that this is
comparable to a cluster a few times the mass of Trapezium, rather than
something particularly massive.

SFR 12 is a more extended region outlined by a shell morphology its
western edge. This bright PAH emission is associated with 2 similar
\hii region velocities. The overall size and shape of the region is
suggested by the shell-like structure visible in the southeastern
edge, but additional \hii velocity measurements are required to ensure
this structure is associated with the known \hii regions.

\subsection{G30 including the W43 region }
\label{G30}
While G30 is also a large region ($\sim 4^{\circ}$ along the major
axis), it is reasonably simple. We identify 6 star forming regions
within this WMAP source, SFR 13-18, but two are clearly dominant; SFR
17 associated with W43, and SFR 14.

SFR 17 consists of a bright central region with a shell outlined by a
series of \hii regions all at similar velocities. W43, located at
($l$,$b$) = (30.75$^\circ$, -0.02$^\circ$), lies inside SFR 17. This
central region has been identified as a ministarburst region with the
detection of a cluster of Wolf-Rayet and OB stars, undergoing a second
generation of star formation through the submillimeter observations of
massive young stellar objects \citep{motte03, lester85, blum99}. We
identify a much larger shell surrounding this region at similar radial
velocities to the central complex. There is no obvious bubble
associated with this SFR.

SFR 16, at ($l$,$b$) = (30.54$^\circ$, 0.02$^\circ$), is contained
completely inside and is much smaller than SFR 17. However, SFR 16 is
distinguished by a significantly different radial velocity (44 vs. 99
km s$^{-1}$).

SFR 14 is located at a similar velocity to SFR 17, but is separated on
the sky by an angle of $\sim0.6^{\circ}$ and it is located at the far
distance ($8.7 \kpc$) while SFR 17 is located at near distance ($6.3
\kpc$). SFR 14 resides in the same 8 micron background as SFR 17. It
is known to be forming stars, as it contains a hot molecular core and
is associated with a molecular cloud \citep{olmi03, pratap99}.

The remainder of the star forming regions within this WMAP source are
low brightness sources in PAH emission and are not expected to
contribute significantly to the total young stellar population.

\subsection{G34}
 \label{G34}
We identify 3 well separated star forming regions in G34, SFRs 19, 20
and 21. The dominant region is SFR 19, coincident with the majority of
the background PAH emission in this area. At the heart of this star
forming region is the cometary \hii region G34.3+0.2
\citep{reid85}. Outside the inner 10 arcminutes of the radio source,
we find three bubble structures (identified as bubbles 22, 23 and 24),
indicating that this region is composed of more than just the single
O7V star assumed by \citet{rr02}.

SFR 21 is suggested primarily by its morphology, as only two \hii
region velocities are identified. The most prominent structure is on
the east, where we identify a number of bubble-like objects. Our
primary motivation in associating this large diffuse region together
is the existence of structure all along the perimeter of the bubble,
following the shell of the structure on the east. It is also possible
that this region consists of two separate star forming regions, on the
east and the west, but in that case it is unclear where the boundary
between the two would be.

In Paper I, this WMAP source was assigned a distance of $10.5 \kpc$
taken from \citet{russeil03}; this is consistent with our present
estimate to the distance to SFR 21. However, from the PAH emission
evident in the Spitzer image, the majority of the free-free emission
may well be coming from SFR 19, located at a distance of only $2.2
\kpc$.  This would reduce the ionizing luminosity of the region
dramatically.

\subsection{G37 including the W47 region }
\label{G37}
Two star forming regions have been identified in this WMAP region,
both at similar velocities and exhibiting ring structure. SFR 22,
which includes W47, is the larger source. There are 14 known \hii
regions, located in the upper half of the bubble. While the southern
shell is more diffuse, we do not classify this bubble as blown out, as
a wall is still visible in the PAH emission. Much diffuse PAH emission
is visible outside this region but specifically towards the brighter
regions on the northern shell, implying that the majority of the
emission is due to the triggered regions on the shell wall rather than
a central source. This suggests that the region is old. The uneven
illumination of the shell structure further supports this
assessment. W47, located on the northeastern shell of SFR 22, has not
been extensively studied in the literature.

SFR 23 has similar radial velocities, but we are unaware of any
resolution of the kinematic distance ambiguity in this case.  A nearly
circular shell structure is visible with a gap on the southeastern
wall.

\subsection{G49 including the W51 region }
\label{G49}
The region containing WMAP source G49 is a well studied area owing to
the presence of W51. We identify two star forming regions within this
source. The chaotic structure in the PAH emission indicates that these
regions are particularly active and young. This chaotic structure
makes it difficult to determine the actual shape and size of the star
forming regions. In the case of SFR 25, the shape is motivated by
shell structure on the northwestern edge. For SFR 24, the shape is
motivated by the distribution of \hii regions.  In both cases, we
suspect that the true extent of the star forming regions may be
larger. The distances to these regions are similar, 5.6 kpc for SFR 24
and 5.7 kpc for SFR 25. These distances are consistent with the
distance of 5.1 kpc determined through trigonometric parallax to W 51
IRS2 \citep{xu09}. Ultimately, we have separated these regions into
two due to the separation of known \hii regions into two clusters on
the sky. The two regions have previously been separated in the
literature, where SFR 24 is labelled W51B and SFR 25 is W51A.

W51 is a well known star formation region
\citep{kang09}. \citet{conti04} separate SFR 25 into W51A and W51
West, placing them at a similar distance of 5.1 kpc. They determine an
ionizing luminosity for both these sources of $\textrm{Q}_{0} = 1.01
\times 10^{51}$ from ground based observations of the free-free
flux. For SFR 24, they determine an ionizing luminosity of
$\textrm{Q}_{0} = 1.07 \times 10^{50}$. Adjusting for the difference
in the presumed distance, the total ionizing luminosity from the three
Conti \& Crowther sources accounts for a quarter of the ionizing
luminosity determined from the entire WMAP source. This is not
surprising; as we indicated in Paper I, the bulk of the ionizing
luminosity leaks beyond the more compact \hii region defined by
ground-based observations into the extended low density (ELD) region.

\citet{fig08} propose a smaller distance of 2.0 kpc to W51A based on
the spectrophotometric distance of a number of O stars observed in the
region, in contrast to the extensive kinematic and now trigonometric
parallax distances determined. In the spectral classification of their
target stars, there were unable to determine the luminosity class and
argue for ZAMS based on the presence of ongoing star formation
indicating a young region. However we have determined the dynamical
age of the system to be two million years, which is sufficient for the
most massive of stars to evolve off the main sequence. Further, the
difference in the smaller and larger distances corresponds to a
difference of 2 in magnitude, which in turn corresponds reasonably to
the difference between the K-band magnitudes of a mid-to-late O star
from luminosity class V to I \citep{martins06}. Thus a more consistent
story is that the identified stars are actually supergiants at the
original distance of the source, 5.1 kpc.

\subsection{G283 including the NGC 3199, RCW 49 and Westerlund 2
  regions}

We identify one star forming region, SFR 26, in WMAP source G283.
Unlike the regions closer to the Galactic centre, there is little
confusion in this source. In fact, SFR 26 is similar in shape and size
to the WMAP source. We use the MSX image in this region as only the
eastern edge of this source has been included in the GLIMPSE survey.

SFR 26 is associated with NGC 3199 and RCW 49, both known regions of
active star formation. NGC 3199, located towards the southwest of the
WMAP source, is associated with the Wolf-Rayet star WR 18
\citep{deha74}. We know of no in-depth attempt to find a cluster in
this region.

RCW 49 is located in the northeast half of the WMAP region and is
associated with the brightest patch of PAH emission within the WMAP
source. RCW 49 is a well known massive star forming region with its
embedded massive cluster Westerlund 2. For a review of this
association, see \S 2 of \citet{tsu07}. Extensive surveys of the
cluster have been completed in the infrared \citep{ascenso07}, and
x-rays \citep{tsu07}. \citet{ascenso07} find a total cluster mass of
$7\times10^{3}$ M$_{\sun}$ using a distance of $2.8 \kpc$ as compared
to our distance of $4 \kpc$. This is a factor of four smaller than the
estimate of the total stellar mass that we determine for the entire
WMAP source using their distance.  This indicates that a substantial
number of stars associated with Westerlund 2 have evaded detection
thus far and/or there is more than one substantial cluster associated
with this star forming region.
  
\subsection{G291 including the NGC 3603 and NGC 3576 regions}

We find two overlapping star forming regions associated with the WMAP
source found at G291; SFR 27 and 28. The separation of these two
regions is motivated by the substantial difference in \hii
velocities. In SFR 27, the mean velocity is $-22 \kms$ and extends to
$-16 \kms$ while SFR 28 has a mean velocity of $16 \kms$ and extends
to $9 \kms$.  Note that the bright region associated with NGC 3576 at
($l$,$b$) = (291.3$^\circ$, -0.7$^\circ$) in the shell of SFR 27 is
projected to lie within SFR 28.

\citet{persi94} find a young massive cluster embedded in NGC 3576,
detecting 40 cluster members with K band photometry. \citet{barbosa03}
find a number of massive YSOs, indicating active star formation. NGC
3576 is located on the shell of the star forming region, suggesting
that it may have been induced by a possible central cluster. This is
consistent with the small size of the NGC 3576 region, indicating its
youth.

SFR 28 is dominated by NGC 3603, located centrally within the region.
Morphologically, it appears that material being swept up in the
northwestern edge of the shell is triggering further star formation.
This appears to also be the case at the location ($l$,$b$) =
(291.1$^\circ$, -0.75$^\circ$), which happens to overlap SFR 27. NGC
3603 is a known massive star forming region powered by HD 97950, often
labelled a ``starburst cluster''\citep[for an extensive review
see][]{melena08}. \citet{conti04} determine a total ionizing
luminosity from NGC 3576 and NGC 3603 of $\textrm{Q}_{0} = 3.4\times
10^{51}$, similar to the ionizing luminosity for the entire WMAP
source, $\textrm{Q}_{0} = 4.2\times 10^{51}$.

\subsection{G298}

G298 is an archetypal region of triggered star formation. We identify
a single star forming region associated with this WMAP source, SFR 29.
In this star forming region, we see an evacuated bubble with a
continuous shell in the PAH emission, with elephant trunk structures
and three possible locations of triggered star formation on the rim of
the bubble; on the northeast, northwest, and southwest.  A great deal
of PAH emission is coming from north of the source, consistent with UV
photons leaking out of the central bubble and illuminating further
material, with the majority of material being closer to the galactic
midplane.

\citet{conti04} identify only the northeast and northwest shell
regions which they identify at ($l$,$b$) = (298.227$^\circ$,
-0.340$^\circ$) and ($l$,$b$) = (298.862$^\circ$, -0.438$^\circ$),
with a total ionizing luminosity of $\textrm{Q}_{0} = 1.5\times
10^{51}$, which is substantially smaller than the value we determine
for the entire WMAP source, $\textrm{Q}_{0} = 7.7\times 10^{51}$.
This larger ionizing luminosity implies that a truly massive cluster
exists in this region, substantially larger than the two sources on
the shell.

\subsection{G311}

The WMAP source G311 is a particularly large on the sky, stretching
$\sim 3$ degrees along the Galactic plane, with a substantial PAH
emission background. As this is a southern galactic plane source, it
is not well studied, and we only identify three star forming regions
convincingly, one of which appears to be dominating this area in the
sky.

SFR 31 is central to both the WMAP source and the background PAH
emission and has a large angular size (nearly $0.5^{\circ}$). Further
it is outlined through its shell morphology as well as the presence of
multiple known \hii regions along the shell. For these reasons, we
infer that it is the dominant star forming region within this WMAP
source. SFR 30 is a much smaller bubble with a similar radial velocity
measurement to SFR 31. However, the resolution of the kinematic
distance ambiguity places SFR 30 at $3.5 \kpc$ and SFR 31 at $7.4
\kpc$.

SFR 32 overlaps the southeastern edge of SFR 31, but is defined by
\hii regions having significantly different radial velocities, placing
it much further away (13.6 kpc versus 7.4 kpc).

The image shows a ridge in the PAH emission continuing from the
southeast wall of SFR 31 strongly suggesting that it is associated
with SFR 31. It is located from ($l$,$b$) = (311.9$^\circ$,
-0.2$^\circ$) to ($l$,$b$) = (312.5$^\circ$, 0.1$^\circ$). However,
there are no measured \hii velocities for this feature, so we cannot
confirm the association.

\subsection{G327}

All the \hii region velocities in G327 lie within $\pm 13 \kms$ of
each other, so we identify only one star forming region, SFR 33. The
morphology revealed by the PAH emission is rather complicated
consisting of multiple bubbles. We interpret this region as an example
of cascading star formation; an instance where multiple generations of
triggered star formation occur in sequence. We find a large, diffuse
shell, outlined by the red ellipse in Figure \ref{G327glimpse}, with a
number of known \hii regions along the wall. Towards the south and the
southwest of the shell, there are two additional clear bubble-like
structures, identified as bubble 39 and 38 in Table
\ref{bubblelist}. These both have additional \hii regions visible on
their walls. We interpret the largest region as a shell produced by an
older central cluster. We speculate that the two smaller (younger?)
bubbles were formed from clusters that were triggered by the original
central cluster. These, in turn, are now triggering a third generation
of star formation. 

A previously identified source, G327.3-0.6 \citep{wyr06}, lies at the
southern edge of SFR 33. This author identifies \hii regions, hot
cores and cold clumps. There is evidence that G327.3-0.6 was triggered by
the expansion of the bubble on the south rim of SFR 33 \citep{minier09}.

We suggest that an older massive cluster should exist towards the
centre of SFR 33 and may be visible in infrared surveys, as this
region is not particularly distant (3.7 kpc).

\subsection{G332}

G332 is another large WMAP source ($\sim 3$ degrees) with only two
distinct clusters in \hii radial velocities. SFR 34 is the
dominant star forming region covering the majority of the WMAP source,
while SFR 35 is a smaller region projected to lie in the interior of 
SFR 34.

The Spitzer image of SFR 34 is dominated by PAH-bright regions along
the southeastern edge. The bright southeastern ridge includes RCW 106
right at the southern edge of the region to the region at ($l$,$b$) =
(333.6$^\circ$, -0.2$^\circ$).

\subsection{G337}

G337 is the widest WMAP source, spanning over 5 degrees in the
Galactic plane. Five star forming regions are identified, SFR
35-40. All of the high surface brightness PAH emission in the Spitzer
image lie within one of our star forming regions. There is a very
prominent background diffuse PAH emission which is difficult to
associate with a specific SFR. Each of the star forming regions are
well separated both on the sky and by radial
velocities. SFR 36, 38 and 40 lack strong central PAH emission suggesting that
they have evacuated their surroundings. In contrast, SFR 37 and 39 show 
filled morphologies, suggestive of youth.

\section{Discussion }
\label{discussion}
It has been noted before that regions with elevated free-free emission
are associated with elevated PAH emission \citep{cohen01, paperI}. In
Paper I, we used this correlation to argue that, based on the higher
resolution 8 $\micronm$ images, both the free-free and PAH emission
are powered by a central source. In this paper, we have shown that
bubbles are associated with all of the sources we have examined. The
bubbles are identified by their bright rims as seen in PAH emission.

In addition, the bubbles are surrounded by an elevated background of
PAH emission. The total luminosity in the $8\micronm$ band of our WMAP
sources is dominated by the elevated background as opposed to the
high-surface brightness emission from the bubble walls. We conclude
that the bulk of the ionizing photons from the central source escape
through the bubble walls to reach distances of 100-200 $\pc$ or more,
confirming the results of Paper I. 

Previous estimates of stellar masses associated with giant \hii
regions have been underestimates of the total stellar mass at those
locations. This follows from the fact that the \hii regions reprocess
only a small fraction of the ionizing photons in a given WMAP
source. 

We suggest that massive clusters inhabit the central regions of the
bubbles. These clusters have evaded detection due to the lack of high
surface brightness \hii or radio continuum emission. We suggest that
this results from a lack of gas and PAH particles in the immediate
vicinity of the young clusters. The clusters have evacuated their
surroundings, leaving little material to reprocess the ionizing
photons emitted by the cluster.

\subsection{Scale Height of O Stars }
\label{scaleheight}
As discussed above, the WMAP sources studied in this paper are
expected to contain approximately one-third of the O stars in the
Galaxy. Using those objects in our catalogue with unique distances, we
estimate the scale height of the O stars. As a first approximation, we
assume that each of the star forming regions contain a similar number
of O stars.  Using this assumption, we construct the cumulative
distribution function of star forming regions as a function of their
height above the galactic plane. The scale height of the 
distribution is

\begin{equation}
  h_{\textrm{*}} = 35\pm 5 \textrm{pc.}
\end{equation}

This value is consistent with that obtained by \citet{elias06} for the
local galactic disk (within 1 kpc) of $h_{\textrm{LGD}}= 34 \pm 3$ pc
for their O-B2 subsample. Both are smaller than the value obtained by
\citet{reed00} of $h_{\textrm{LGD}}= 45 \pm 20$ pc for all OB stars
within a distance of 4 kpc, but easily fall within their error
bars. We note that the scale height of the molecular gas is $h \sim
40$ pc \citep{malhotra94}. 

\subsection{Free-Free to PAH Emission Relationship }
\label{pah}
\begin{figure}
  \includegraphics[width=7in]{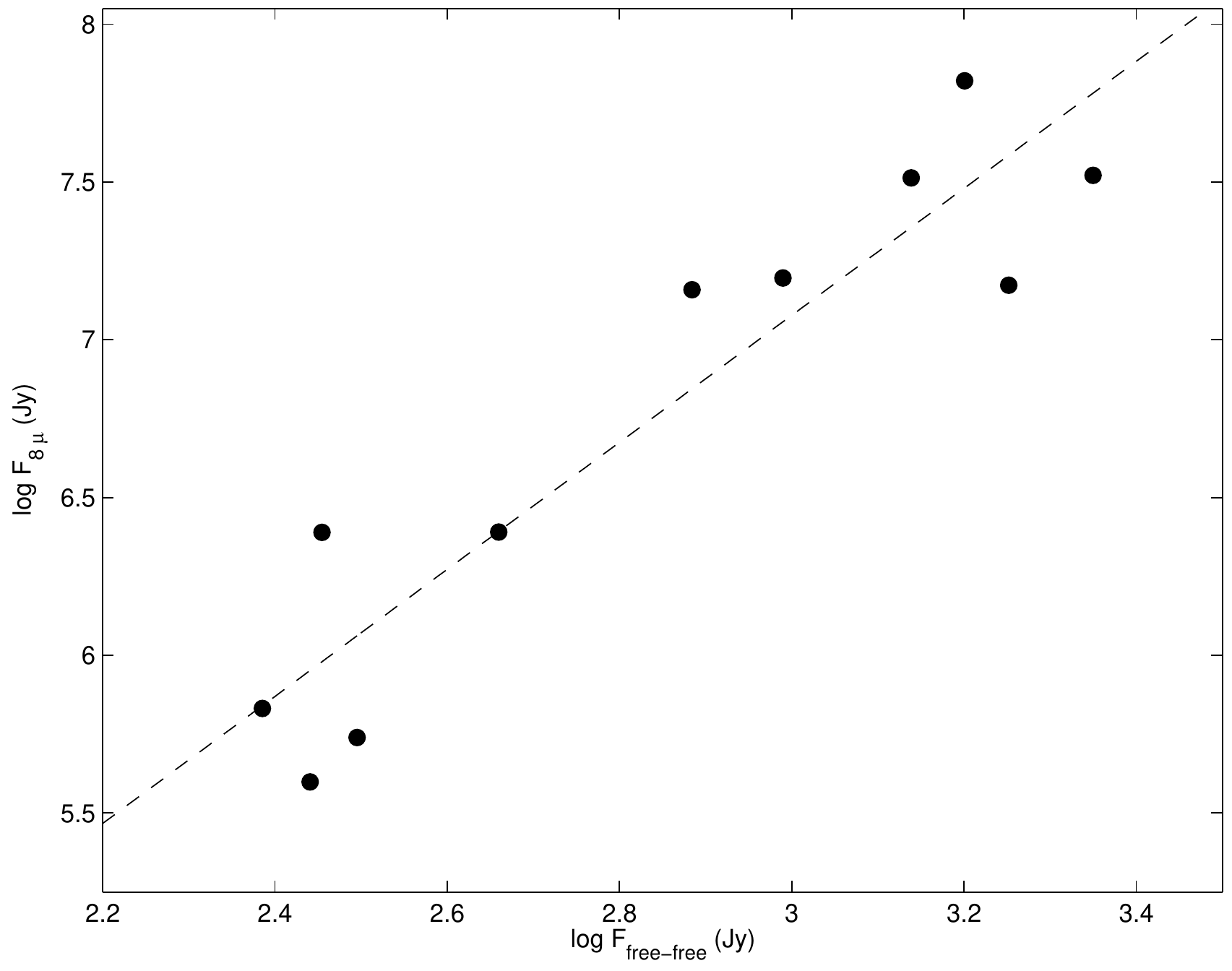}
  \caption{The comparison between the total flux integrated over the
regions presented in Table \ref{wmap} in free-free emission through 
the WMAP free-free foreground emission map at 90 GHz, and in PAH emission
taken through the GLIMPSE 8 micron mosaics. The slope of the best fit 
line is $2.0\pm0.3$. Regions without GLIMPSE coverage are excluded from this plot.  
\label{fluxfigure}}
\end{figure}

A morphological correlation between the MSX 8 micron emission and the
radio continuum \citep{cohen01}. We investigate this correlation using
the WMAP free-free emission maps and the PAH emission from the GLIMPSE
8 micron mosaics.  For each WMAP source, we summed the free-free
emission from the inside the Source Extractor ellipse from Paper I. We
summed the $8\micronm$ from the same ellipse in the Spitzer GLIMPSE
mosaic. The result is shown in Figure \ref{fluxfigure}. From a
least-square fit, we find the following relationship:

\begin{equation}
  F_{\textrm{PAH}}\propto F_{\textrm{ff}}^{2.0\pm0.34} \label{obsrel}
\end{equation}

where $F_{\textrm{PAH}}$ is the integrated 8 micron GLIMPSE flux and
$F_{\textrm{ff}}$ is the integrated 90 GHz free-free emission from the
WMAP free-free foreground emission map. We find a similar
relationship using the MSX Band A integrated flux in place of the
GLIMPSE measurements.

\subsection{Velocity Dispersion of the Molecular Gas}

Both molecular and atomic gas in the disk of the Milky Way are seen to
have supersonic velocity dispersions.  These dispersions are normally
interpreted as being due to turbulence, although their origin is
uncertain. If the motions are due to turbulence, they must be driven,
since undriven turbulence decays on roughly a dynamical time
\citep{maclow98}. Furthermore, the turbulence must be driven on $100
\pc$ scales, since three dimensional turbulence cascades from large
scales to small scales and not the other way around. A number of
driving mechanisms have been proposed \citep{maclow04, miesch94},
including supernovae, stellar winds, and gravitational instabilities,
with no conclusive evidence for any particular mechanism. We
investigate the kinetic energy that is injected into the ISM by the
expansion of the massive bubbles that we identify. We note that all of
the star forming regions we have identified are likely to be in giant
molecular clouds---a preliminary search shows that more than thirty
are in fact in molecular clouds.

We calculate the mechanical luminosity in the expansion of each of
the star forming regions using
\begin{equation}
  L_{mech} = \frac{\pi}{2}\Sigma_{0} \Delta v_{c}^{3} r,
\end{equation}
where $\Sigma_{0} = 170 M_{\sun}$ pc$^{-2}$ \citep{solomon87} is the
surface density of a GMC, and $r$ and $\Delta v_{c}$ are taken from
Table \ref{sfrlist}. For SFRs where the half-spread velocity was not
measured, we used the mean half-spread velocity, $\Delta v$ = 12 km
s$^{-1}$, of the known sources. For cases where the kinematic distance
remains ambiguous, we take the mean radius of the region determined
from both the near and far distances. Since the ratio of the two
distances is less than two, this introduces an error of less than a
factor of two for 9 out of our 40 SFRs. For each of the SFRs, we present
the calculated dynamical properties in Table \ref{dynprop} with the
columns as follows: column (1) the catalogue number, column (2) the
mass swept up in the shell in solar masses, column (3) the kinetic
energy of the swept up shell in erg, and column (4) the mechanical
luminosity produced by the motion of the shell in $\textrm{erg
  s}^{-1}$.  

The total mechanical luminosity being injected into the ISM inside the
solar radius due to the expansion of these SFRs is the sum of the
individual luminosities.  We find
\begin{equation}
  L_{mech} \approx 6.7 \times 10^{38} \left(\frac{\Sigma_{0}}{170 M_{\sun}\pc^{-2}}\right) \textrm{ erg s}^{-1}.
\end{equation} 
This sum encompases only one-third of the star formation in the Galaxy. The other
two-thirds should supply a proportionate amount of mechanical
luminosity, for a total kinetic luminosity of $\sim2\times10^{39}\erg\s^{-1}$.

We compare this luminosity to that required to maintain the velocity
dispersion in the molecular gas within the the solar circle:

\begin{equation}
  L_{turb} \equiv \frac{1}{2}\frac{ M v^3}{2h}, 
\end{equation} 
where $h = 40$ pc is the scale height of the molecular disk, the
molecular gas mass inside the solar circle $M = 1.0 \times 10^{9}
\textrm{M}_{\sun}$, and a molecular gas velocity $v = \sqrt{2 \ln{2}}
\sigma_{mol}$ with $\sigma_{mol} = 7\kms$ \citep{malhotra94}.  The
turbulent luminosity is
\begin{equation}
  L_{turb} \approx 2.4 \times 10^{39} \left(\frac{v}{8 \kms} \right)^3 \textrm{ erg s}^{-1}.
\end{equation} 

We conclude that the mechanical luminosity we see in the bubbles,
multiplied by a factor of three to account for the other two thirds
of Galactic star formation, is sufficient to power the turbulent
luminosity seen in the molecular gas.

We note that the kinetic energies calculated for most of the bubbles
(refer to Table \ref{dynprop}) range from $10^{48}$ to just below
$10^{52}\textrm{ erg s}^{-1}$. The kinetic energies of our most
energetic bubbles are similar to those measured in the superbubbles
identified by \citet{heiles79}, which range from
$4\times10^{51}\erg\s^{-1}$ to $2\times10^{53}\erg\,\s^{-1}$. 

We note that our bubbles are selected by their ionizing luminosity,
i.e., we require a very young stellar population. In addition, we do
not see significant synchrotron radiation from these areas, indicating
the absence of supernova remnants. This implies that the clusters are
too young to have had more than a few supernova explode, and many of
our sources likely have not had any supernovae go off. We infer that
the superbubbles in the Galaxy are not initially driven by the energy
from supernovae, but rather by the energy injected into the medium by
the massive stars during their lifetime. It may well be, however, that
supernovae contribute significantly to the kinetic energy later in the
evolution of a superbubble.

\subsection{Bubbles: Spherical Shells or Flattened Rings?}

The three-dimensional geometry of the bubbles identified by
\citet{churchwell06} has been contested recently. \citet{beaumont10}
propose that Churchwell et al.'s bubbles are flattened
rings. \citet{beaumont10} suggest that the aspect ratios of these
rings may be anywhere from a few to as much as 10. Nineteen of our
forty eight bubbles are in the GLIMPSE bubble catalogs. We test the
assertion of \citet{beaumont10} statistically using the bubbles in
Table \ref{bubblelist}.

We model our bubbles as ellipsoids, following the procedure of
\citet{noumeir99} with semi-axes $a, b, c$ aligned along the x, y and
z axes such that $a \ge b \ge c$. The matrix equation for the
ellipsoid is
\begin{equation}
u^{\intercal} X u = 1 
\end{equation}
where $u$ is an arbitrary position vector. The matrix $X$ is a diagonal
matrix with entries $a^2, b^2, c^2$, corresponding to the semi-axes of
the ellipsoid. We rotate the ellipsoid along the z-axis by an angle
$\theta$, and along the y-axis by an angle $\phi$, producing the
matrix

\begin{equation}
  X' = 
  \left[ \begin{array}{ccc}
      {\scriptstyle \alpha \cos^2 \phi + c^2 \sin^2 \phi } & 
      {\scriptstyle \beta \cos \phi \cos \theta \sin \theta }&
      {\scriptstyle (c^2 - \alpha)\sin \phi \cos \phi} \\
      {\scriptstyle \beta \cos \phi \cos \theta \sin \theta }& 
      {\scriptstyle a^2 \sin^2 \theta + b^2 \cos^2 \theta }&
      {\scriptstyle -\beta \sin \phi \cos \theta \sin \theta }\\
      {\scriptstyle -(\alpha + c^2) \sin \phi \cos \phi }& 
      {\scriptstyle -\beta \sin \phi \cos \theta \sin \theta }& 
      {\scriptstyle \alpha \sin^2 \phi + c^2 \cos^2 \phi}
  \end{array}
\right]
\end{equation}
where we define $\alpha = a^2 \cos^2 \theta + b^2 \sin^2 \theta $, and
$\beta = a^2 - b^2 $.

To determine the observed ellipse on the sky resulting from the
rotation of the ellipsoid, we produce an orthographic projection of
the ellipsoid onto the y-z plane. The resulting projection is
\begin{equation}
  X_P = 
  \left[ \begin{array}{cc}
      a^2 \sin^2 \theta + b^2 \cos^2 \theta & -\beta \sin \phi \cos \theta \sin \theta \\
      -\beta \sin \phi \cos \theta \sin \theta & \alpha \sin^2 \phi + c^2 \cos^2 \phi\\
  \end{array}
\right]
\end{equation}

To minimize the observed axis ratio, we assume the bubbles are
circular rings with $a = b$, implying the semi-major axis of the
projected bubble is $a$, and the projected semi-minor axis is $\sqrt{a^2 \sin^2
  \phi + c^2 \cos^2 \phi}$. The resulting axis ratio, a function of
only the rotation along the y-axis, is given as
\begin{equation} \label{eqn: aspect}
R_{ax}(\phi) = \frac{a}{\sqrt{a^2 \sin^2 \phi + c^2 \cos^2 \phi}}
\end{equation}

We compute the expectation value of the axis ratio $\langle
R_{ax}(a/c)\rangle$, for $a/c=10$ and $a/c=4$. Using eqn. (\ref{eqn:
  aspect}) we find $\langle R_{ax}(10)\rangle = 2.4$ and $\langle
R_{ax}(4)\rangle = 1.8$.  We compare this to our sample of $48$
bubbles, for which we find $\langle R_{ax}\rangle = 1.3\pm0.3$, suggesting
that the intrinsic aspect ratio $a/c\lesssim4$ at the two sigma
level. In fact $\langle R_{ax}\rangle = 1.3$ corresponds to a mean
intrinsic aspect ratio $a/c=1.7$.

A more sensitive test is that of the maximum aspect ratio; the maximum
observed aspect ratio is 2.2, corresponding to a maximum inclination
angle of $31^{\circ}$ from the face-on position of a ring with $a/c=4$. For a
simple random distribution of inclination angles, we would expect
$48\%$, or $23\pm7$ of the bubbles to have an aspect ratio greater
than $2.2$.

We note that if the observed maximum aspect ratio ($2.2$) is in fact
the true mean aspect ratio, then $\langle R_{ax}(2.2)\rangle=1.4$,
consistent with the observed mean aspect ratio.

We find an insufficient number of observed bubbles with large aspect
ratios to support the geometry of a flattened ring and the original
picture of nearly-spherical shells is more likely to be physically
correct.

\section{Summary }
\label{summary}
From an analysis of the 13 most luminous sources found in the WMAP
free-free emission map, we have identified 40 star forming
regions using a combination of morphology in PAH emission from the
Spitzer GLIMPSE and MSX surveys and velocities of known \hii regions
at the same location on the sky. For all 31 of the 40 regions we
are able to determine a unique distance associated with the radial
velocity measurements, allowing us to determine the physical
size of these regions. For 20 of these star forming regions, there are
a sufficient number of \hii region velocity measurements to estimate the
expansion speed of the region and consequently, a dynamical age.

We catalogue 48 bubble-like objects, 19 of which have been previously
catalogued by \citet{churchwell06}. For most of these bubbles, we were
able to determine a physical size. We also present a comprehensive
list of the radio recombination line velocity measurements associated
with each star forming region.

We discuss each of the WMAP sources individually, commenting on the
morphology within the source. Many of the northern sky star forming
regions are associated with previously identified star forming
regions. In most cases, the previous masses (or luminosities) for
embedded clusters are substantially underestimated, as earlier studies
often did not take into account the leakage of ionizing photons from
the central ionizing source.  We also find substantial evidence of
triggered star formation in many of the star forming regions. A number
of our star forming regions, especially the southern sky regions, have
not been previously investigated or identified as possible locations
of massive stellar clusters.

We determine the scale height of the Galactic O stars to be
$h_{\textrm{*}} = 35 \pm 5 \pc$, consistent with previous values for
the scale height of OB stars in the local galactic plane.  We
determine a strong empirical relationship between the total integrated
PAH and free-free emission; $F_{\textrm{PAH}} \propto
F^2_{\textrm{ff}}$.

We show that the kinetic energy injected per unit time into the
galactic medium by the expansion of the star forming regions is
similar to that required to maintain the velocity dispersion seen
in the Galactic molecular gas. Thus we propose that the expansion
of the star forming regions is a primary driver of turbulence in the
molecular gas of the Galaxy. We further show that the kinetic energy
of the identified bubbles is similar to that of known superbubbles,
implying that superbubbles are being produced before any supernova
activity has taken place within the cluster.

Finally, we show that the three dimensional geometry of the bubbles is
more consistent with a nearly-spherical shell rather than a flattened
ring geometry based on statistics of the identified sample.

\acknowledgements

We have benefited from greatly from ongoing discussions with
C. Matzner, D-S. Moon, P. G. Martin, and R. Breton. This work is based
[in part] on observations made with the Spitzer Space Telescope, which
is operated by the Jet Propulsion Laboratory, California Institute of
Technology under a contract with NASA.  This research has made use of
the SIMBAD database, operated at CDS, Strasbourg, France This research
made use of data products from the Midcourse Space Experiment.
Processing of the data was funded by the Ballistic Missile Defense
Organization with additional support from NASA Office of Space
Science.  This research has also made use of the NASA/ IPAC Infrared
Science Archive, which is operated by the Jet Propulsion Laboratory,
California Institute of Technology, under contract with the National
Aeronautics and Space Administration.  This research made use of
Montage, funded by the National Aeronautics and Space Administration's
Earth Science Technology Office, Computation Technologies Project,
under Cooperative Agreement Number NCC5-626 between NASA and the
California Institute of Technology.  Montage is maintained by the
NASA/IPAC Infrared Science Archive. N.M. is supported in part by the
Canada Research Chair program and by NSERC of Canada.

\bibliography{paperii}

\begin{thebibliography}{88}
\expandafter\ifx\csname natexlab\endcsname\relax\def\natexlab#1{#1}\fi

\bibitem[{{Anantharamaiah}(1985{\natexlab{a}})}]{anan85}
{Anantharamaiah}, K.~R. 1985{\natexlab{a}}, Journal of Astrophysics and
  Astronomy, 6, 177

\bibitem[{{Anantharamaiah}(1985{\natexlab{b}})}]{anan85a}
---. 1985{\natexlab{b}}, Journal of Astrophysics and Astronomy, 6, 203

\bibitem[{{Ascenso} {et~al.}(2007){Ascenso}, {Alves}, {Beletsky}, \&
  {Lago}}]{ascenso07}
{Ascenso}, J., {Alves}, J., {Beletsky}, Y., \& {Lago}, M.~T.~V.~T. 2007, \aap,
  466, 137

\bibitem[{{Barbosa} {et~al.}(2003){Barbosa}, {Damineli}, {Blum}, \&
  {Conti}}]{barbosa03}
{Barbosa}, C.~L., {Damineli}, A., {Blum}, R.~D., \& {Conti}, P.~S. 2003, \aj,
  126, 2411

\bibitem[{{Beaumont} \& {Williams}(2010)}]{beaumont10}
{Beaumont}, C.~N., \& {Williams}, J.~P. 2010, \apj, 709, 791

\bibitem[{{Benjamin} {et~al.}(2003){Benjamin}, {Churchwell}, {Babler}, {Bania},
  {Clemens}, {Cohen}, {Dickey}, {Indebetouw}, {Jackson}, {Kobulnicky},
  {Lazarian}, {Marston}, {Mathis}, {Meade}, {Seager}, {Stolovy}, {Watson},
  {Whitney}, {Wolff}, \& {Wolfire}}]{benjamin03}
{Benjamin}, R.~A., {et~al.} 2003, \pasp, 115, 953

\bibitem[{{Bennett} {et~al.}(2003{\natexlab{a}}){Bennett}, {Hill}, {Hinshaw},
  {Nolta}, {Odegard}, {Page}, {Spergel}, {Weiland}, {Wright}, {Halpern},
  {Jarosik}, {Kogut}, {Limon}, {Meyer}, {Tucker}, \& {Wollack}}]{bennett03}
{Bennett}, C.~L., {et~al.} 2003{\natexlab{a}}, \apjs, 148, 97

\bibitem[{{Bennett} {et~al.}(2003{\natexlab{b}}){Bennett}, {Hill}, {Hinshaw},
  {Nolta}, {Odegard}, {Page}, {Spergel}, {Weiland}, {Wright}, {Halpern},
  {Jarosik}, {Kogut}, {Limon}, {Meyer}, {Tucker}, \& {Wollack}}]{wmappaper}
---. 2003{\natexlab{b}}, \apjs, 148, 97

\bibitem[{{Blum} {et~al.}(2000){Blum}, {Conti}, \& {Damineli}}]{blum00}
{Blum}, R.~D., {Conti}, P.~S., \& {Damineli}, A. 2000, \aj, 119, 1860

\bibitem[{{Blum} {et~al.}(1999){Blum}, {Damineli}, \& {Conti}}]{blum99}
{Blum}, R.~D., {Damineli}, A., \& {Conti}, P.~S. 1999, \aj, 117, 1392

\bibitem[{{Bronfman} {et~al.}(1996){Bronfman}, {Nyman}, \& {May}}]{bronfman96}
{Bronfman}, L., {Nyman}, L., \& {May}, J. 1996, \aaps, 115, 81

\bibitem[{{Caswell}(1972)}]{caswell72}
{Caswell}, J.~L. 1972, Australian Journal of Physics, 25, 443

\bibitem[{{Caswell} \& {Haynes}(1987)}]{caswell87}
{Caswell}, J.~L., \& {Haynes}, R.~F. 1987, \aap, 171, 261

\bibitem[{{Caswell} {et~al.}(1975){Caswell}, {Murray}, {Roger}, {Cole}, \&
  {Cooke}}]{caswell75}
{Caswell}, J.~L., {Murray}, J.~D., {Roger}, R.~S., {Cole}, D.~J., \& {Cooke},
  D.~J. 1975, \aap, 45, 239

\bibitem[{{Churchwell} {et~al.}(1990){Churchwell}, {Walmsley}, \&
  {Cesaroni}}]{churchwell90}
{Churchwell}, E., {Walmsley}, C.~M., \& {Cesaroni}, R. 1990, \aaps, 83, 119

\bibitem[{{Churchwell} {et~al.}(2006){Churchwell}, {Povich}, {Allen}, {Taylor},
  {Meade}, {Babler}, {Indebetouw}, {Watson}, {Whitney}, {Wolfire}, {Bania},
  {Benjamin}, {Clemens}, {Cohen}, {Cyganowski}, {Jackson}, {Kobulnicky},
  {Mathis}, {Mercer}, {Stolovy}, {Uzpen}, {Watson}, \& {Wolff}}]{churchwell06}
{Churchwell}, E., {et~al.} 2006, \apj, 649, 759

\bibitem[{{Clark} {et~al.}(2005){Clark}, {Negueruela}, {Crowther}, \&
  {Goodwin}}]{clark05}
{Clark}, J.~S., {Negueruela}, I., {Crowther}, P.~A., \& {Goodwin}, S.~P. 2005,
  \aap, 434, 949

\bibitem[{{Clark} {et~al.}(2009){Clark}, {Negueruela}, {Davies}, {Larionov},
  {Ritchie}, {Figer}, {Messineo}, {Crowther}, \& {Arkharov}}]{clark09}
{Clark}, J.~S., {et~al.} 2009, \aap, 498, 109

\bibitem[{{Clemens}(1985)}]{clemens85}
{Clemens}, D.~P. 1985, \apj, 295, 422

\bibitem[{{Cohen} \& {Green}(2001)}]{cohen01}
{Cohen}, M., \& {Green}, A.~J. 2001, \mnras, 325, 531

\bibitem[{{Conti} \& {Crowther}(2004)}]{conti04}
{Conti}, P.~S., \& {Crowther}, P.~A. 2004, \mnras, 355, 899

\bibitem[{{Corbel} \& {Eikenberry}(2004)}]{corbel04}
{Corbel}, S., \& {Eikenberry}, S.~S. 2004, \aap, 419, 191

\bibitem[{{Corbel} {et~al.}(1997){Corbel}, {Wallyn}, {Dame}, {Durouchoux},
  {Mahoney}, {Vilhu}, \& {Grindlay}}]{corbel97}
{Corbel}, S., {Wallyn}, P., {Dame}, T.~M., {Durouchoux}, P., {Mahoney}, W.~A.,
  {Vilhu}, O., \& {Grindlay}, J.~E. 1997, \apj, 478, 624

\bibitem[{{Davies} {et~al.}(2007){Davies}, {Figer}, {Kudritzki}, {MacKenty},
  {Najarro}, \& {Herrero}}]{davies07}
{Davies}, B., {Figer}, D.~F., {Kudritzki}, R., {MacKenty}, J., {Najarro}, F.,
  \& {Herrero}, A. 2007, \apj, 671, 781

\bibitem[{{Deharveng} \& {Maucherat}(1974)}]{deha74}
{Deharveng}, L., \& {Maucherat}, M. 1974, \aap, 34, 465

\bibitem[{{Downes} {et~al.}(1980){Downes}, {Wilson}, {Bieging}, \&
  {Wink}}]{downes80}
{Downes}, D., {Wilson}, T.~L., {Bieging}, J., \& {Wink}, J. 1980, \aaps, 40,
  379

\bibitem[{{Eikenberry} {et~al.}(2004){Eikenberry}, {Matthews}, {LaVine},
  {Garske}, {Hu}, {Jackson}, {Patel}, {Barry}, {Colonno}, {Houck}, {Wilson},
  {Corbel}, \& {Smith}}]{eikenberry04}
{Eikenberry}, S.~S., {et~al.} 2004, \apj, 616, 506

\bibitem[{{Elias} {et~al.}(2006){Elias}, {Cabrera-Ca{\~n}o}, \&
  {Alfaro}}]{elias06}
{Elias}, F., {Cabrera-Ca{\~n}o}, J., \& {Alfaro}, E.~J. 2006, \aj, 131, 2700

\bibitem[{{Elmegreen}(1998)}]{elmegreen98}
{Elmegreen}, B.~G. 1998, in Astronomical Society of the Pacific Conference
  Series, Vol. 148, Origins, ed. {C.~E.~Woodward, J.~M.~Shull, \&
  H.~A.~Thronson Jr.}, 150--+

\bibitem[{{Figer} {et~al.}(1999){Figer}, {Kim}, {Morris}, {Serabyn}, {Rich}, \&
  {McLean}}]{figer99}
{Figer}, D.~F., {Kim}, S.~S., {Morris}, M., {Serabyn}, E., {Rich}, R.~M., \&
  {McLean}, I.~S. 1999, \apj, 525, 750

\bibitem[{{Figer} {et~al.}(2006){Figer}, {MacKenty}, {Robberto}, {Smith},
  {Najarro}, {Kudritzki}, \& {Herrero}}]{figer06}
{Figer}, D.~F., {MacKenty}, J.~W., {Robberto}, M., {Smith}, K., {Najarro}, F.,
  {Kudritzki}, R.~P., \& {Herrero}, A. 2006, \apj, 643, 1166

\bibitem[{{Figuer{\^e}do} {et~al.}(2008){Figuer{\^e}do}, {Blum}, {Damineli},
  {Conti}, \& {Barbosa}}]{fig08}
{Figuer{\^e}do}, E., {Blum}, R.~D., {Damineli}, A., {Conti}, P.~S., \&
  {Barbosa}, C.~L. 2008, \aj, 136, 221

\bibitem[{{Gardner} \& {Whiteoak}(1978)}]{gardner78}
{Gardner}, F.~F., \& {Whiteoak}, J.~B. 1978, \mnras, 183, 711

\bibitem[{{Georgelin} \& {Georgelin}(1970)}]{georgelin70}
{Georgelin}, Y.~P., \& {Georgelin}, Y.~M. 1970, \aap, 7, 133

\bibitem[{{Gold} {et~al.}(2009){Gold}, {Bennett}, {Hill}, {Hinshaw}, {Odegard},
  {Page}, {Spergel}, {Weiland}, {Dunkley}, {Halpern}, {Jarosik}, {Kogut},
  {Komatsu}, {Larson}, {Meyer}, {Nolta}, {Wollack}, \& {Wright}}]{gold09}
{Gold}, B., {et~al.} 2009, \apjs, 180, 265

\bibitem[{{Harper-Clark} \& {Murray}(2009)}]{harperclark09}
{Harper-Clark}, E., \& {Murray}, N. 2009, \apj, 693, 1696

\bibitem[{{Heiles}(1979)}]{heiles79}
{Heiles}, C. 1979, \apj, 229, 533

\bibitem[{{Ho}(1997)}]{ho97}
{Ho}, L.~C. 1997, in Revista Mexicana de Astronomia y Astrofisica Conference
  Series, Vol.~6, Revista Mexicana de Astronomia y Astrofisica Conference
  Series, ed. {J.~Franco, R.~Terlevich, \& A.~Serrano}, 5--+

\bibitem[{{Ikeda} {et~al.}(2001){Ikeda}, {Ohishi}, {Nummelin}, {Dickens},
  {Bergman}, {Hjalmarson}, \& {Irvine}}]{ikeda01}
{Ikeda}, M., {Ohishi}, M., {Nummelin}, A., {Dickens}, J.~E., {Bergman}, P.,
  {Hjalmarson}, {\AA}., \& {Irvine}, W.~M. 2001, \apj, 560, 792

\bibitem[{{Kang} {et~al.}(2009){Kang}, {Bieging}, {Povich}, \& {Lee}}]{kang09}
{Kang}, M., {Bieging}, J.~H., {Povich}, M.~S., \& {Lee}, Y. 2009, \apj, 706, 83

\bibitem[{{Kim} \& {Koo}(2001)}]{kim01}
{Kim}, K., \& {Koo}, B. 2001, \apj, 549, 979

\bibitem[{{Kim} \& {Koo}(2002)}]{kim02}
---. 2002, \apj, 575, 327

\bibitem[{{Kolpak} {et~al.}(2003){Kolpak}, {Jackson}, {Bania}, {Clemens}, \&
  {Dickey}}]{kolpak03}
{Kolpak}, M.~A., {Jackson}, J.~M., {Bania}, T.~M., {Clemens}, D.~P., \&
  {Dickey}, J.~M. 2003, \apj, 582, 756

\bibitem[{{Kuchar} \& {Bania}(1994)}]{kuchar94}
{Kuchar}, T.~A., \& {Bania}, T.~M. 1994, \apj, 436, 117

\bibitem[{{Kulkarni} \& {Frail}(1993)}]{kul93}
{Kulkarni}, S.~R., \& {Frail}, D.~A. 1993, \nat, 365, 33

\bibitem[{{Leahy} \& {Tian}(2008)}]{leahy08}
{Leahy}, D.~A., \& {Tian}, W.~W. 2008, \aj, 135, 167

\bibitem[{{Lester} {et~al.}(1985){Lester}, {Dinerstein}, {Werner}, {Harvey},
  {Evans}, \& {Brown}}]{lester85}
{Lester}, D.~F., {Dinerstein}, H.~L., {Werner}, M.~W., {Harvey}, P.~M.,
  {Evans}, II, N.~J., \& {Brown}, R.~L. 1985, \apj, 296, 565

\bibitem[{{Lockman}(1976)}]{lockman76}
{Lockman}, F.~J. 1976, \apj, 209, 429

\bibitem[{{Lockman}(1989)}]{lockman89}
---. 1989, \apjs, 71, 469

\bibitem[{{Lockman} {et~al.}(1996){Lockman}, {Pisano}, \& {Howard}}]{lockman96}
{Lockman}, F.~J., {Pisano}, D.~J., \& {Howard}, G.~J. 1996, \apj, 472, 173

\bibitem[{{Mac Low} \& {Klessen}(2004)}]{maclow04}
{Mac Low}, M., \& {Klessen}, R.~S. 2004, Reviews of Modern Physics, 76, 125

\bibitem[{{Mac Low} {et~al.}(1998){Mac Low}, {Klessen}, {Burkert}, \&
  {Smith}}]{maclow98}
{Mac Low}, M., {Klessen}, R.~S., {Burkert}, A., \& {Smith}, M.~D. 1998,
  Physical Review Letters, 80, 2754

\bibitem[{{Malhotra}(1994)}]{malhotra94}
{Malhotra}, S. 1994, \apj, 433, 687

\bibitem[{{Martins} \& {Plez}(2006)}]{martins06}
{Martins}, F., \& {Plez}, B. 2006, \aap, 457, 637

\bibitem[{{Melena} {et~al.}(2008){Melena}, {Massey}, {Morrell}, \&
  {Zangari}}]{melena08}
{Melena}, N.~W., {Massey}, P., {Morrell}, N.~I., \& {Zangari}, A.~M. 2008, \aj,
  135, 878

\bibitem[{{Miesch} \& {Bally}(1994)}]{miesch94}
{Miesch}, M.~S., \& {Bally}, J. 1994, \apj, 429, 645

\bibitem[{{Minier} {et~al.}(2009){Minier}, {Andr{\'e}}, {Bergman}, {Motte},
  {Wyrowski}, {Le Pennec}, {Rodriguez}, {Boulade}, {Doumayrou}, {Dubreuil},
  {Gallais}, {Hamon}, {Lagage}, {Lortholary}, {Martignac}, {Rev{\'e}ret},
  {Roussel}, {Talvard}, {Willmann}, \& {Olofsson}}]{minier09}
{Minier}, V., {et~al.} 2009, \aap, 501, L1

\bibitem[{{Motte} {et~al.}(2003){Motte}, {Schilke}, \& {Lis}}]{motte03}
{Motte}, F., {Schilke}, P., \& {Lis}, D.~C. 2003, \apj, 582, 277

\bibitem[{{Murray} \& {Rahman}(2010)}]{paperI}
{Murray}, N., \& {Rahman}, M. 2010, \apj, 709, 424

\bibitem[{{Nakanishi} \& {Sofue}(2006)}]{naka06}
{Nakanishi}, H., \& {Sofue}, Y. 2006, \pasj, 58, 847

\bibitem[{Noumeir(1999)}]{noumeir99}
Noumeir, R. 1999, Pattern Recognition Letters, 20, 585

\bibitem[{{Olmi} {et~al.}(2003){Olmi}, {Cesaroni}, {Hofner}, {Kurtz},
  {Churchwell}, \& {Walmsley}}]{olmi03}
{Olmi}, L., {Cesaroni}, R., {Hofner}, P., {Kurtz}, S., {Churchwell}, E., \&
  {Walmsley}, C.~M. 2003, \aap, 407, 225

\bibitem[{{Palagi} {et~al.}(1993){Palagi}, {Cesaroni}, {Comoretto}, {Felli}, \&
  {Natale}}]{palagi93}
{Palagi}, F., {Cesaroni}, R., {Comoretto}, G., {Felli}, M., \& {Natale}, V.
  1993, \aaps, 101, 153

\bibitem[{{Persi} {et~al.}(1994){Persi}, {Roth}, {Tapia}, {Ferrari-Toniolo}, \&
  {Marenzi}}]{persi94}
{Persi}, P., {Roth}, M., {Tapia}, M., {Ferrari-Toniolo}, M., \& {Marenzi},
  A.~R. 1994, \aap, 282, 474

\bibitem[{{Pestalozzi} {et~al.}(2005){Pestalozzi}, {Minier}, \&
  {Booth}}]{pestalozzi05}
{Pestalozzi}, M.~R., {Minier}, V., \& {Booth}, R.~S. 2005, \aap, 432, 737

\bibitem[{{Pratap} {et~al.}(1999){Pratap}, {Megeath}, \& {Bergin}}]{pratap99}
{Pratap}, P., {Megeath}, S.~T., \& {Bergin}, E.~A. 1999, \apj, 517, 799

\bibitem[{{Price} {et~al.}(2001){Price}, {Egan}, {Carey}, {Mizuno}, \&
  {Kuchar}}]{price01}
{Price}, S.~D., {Egan}, M.~P., {Carey}, S.~J., {Mizuno}, D.~R., \& {Kuchar},
  T.~A. 2001, \aj, 121, 2819

\bibitem[{{Quireza} {et~al.}(2006){Quireza}, {Rood}, {Balser}, \&
  {Bania}}]{quireza06}
{Quireza}, C., {Rood}, R.~T., {Balser}, D.~S., \& {Bania}, T.~M. 2006, \apjs,
  165, 338

\bibitem[{{Reed}(2000)}]{reed00}
{Reed}, B.~C. 2000, \aj, 120, 314

\bibitem[{{Reid} \& {Ho}(1985)}]{reid85}
{Reid}, M.~J., \& {Ho}, P.~T.~P. 1985, \apjl, 288, L17

\bibitem[{{Rodr{\'{\i}}guez-Rico} {et~al.}(2002){Rodr{\'{\i}}guez-Rico},
  {Rodr{\'{\i}}guez}, \& {G{\'o}mez}}]{rr02}
{Rodr{\'{\i}}guez-Rico}, C.~A., {Rodr{\'{\i}}guez}, L.~F., \& {G{\'o}mez}, Y.
  2002, Revista Mexicana de Astronomia y Astrofisica, 38, 3

\bibitem[{{Russeil}(2003)}]{russeil03}
{Russeil}, D. 2003, \aap, 397, 133

\bibitem[{{Russeil} {et~al.}(2005){Russeil}, {Adami}, {Amram}, {Le Coarer},
  {Georgelin}, {Marcelin}, \& {Parker}}]{russeil05}
{Russeil}, D., {Adami}, C., {Amram}, P., {Le Coarer}, E., {Georgelin}, Y.~M.,
  {Marcelin}, M., \& {Parker}, Q. 2005, \aap, 429, 497

\bibitem[{{Sewilo} {et~al.}(2004){Sewilo}, {Watson}, {Araya}, {Churchwell},
  {Hofner}, \& {Kurtz}}]{sewilo04}
{Sewilo}, M., {Watson}, C., {Araya}, E., {Churchwell}, E., {Hofner}, P., \&
  {Kurtz}, S. 2004, \apjs, 154, 553

\bibitem[{{Skrutskie} {et~al.}(2006){Skrutskie}, {Cutri}, {Stiening},
  {Weinberg}, {Schneider}, {Carpenter}, {Beichman}, {Capps}, {Chester},
  {Elias}, {Huchra}, {Liebert}, {Lonsdale}, {Monet}, {Price}, {Seitzer},
  {Jarrett}, {Kirkpatrick}, {Gizis}, {Howard}, {Evans}, {Fowler}, {Fullmer},
  {Hurt}, {Light}, {Kopan}, {Marsh}, {McCallon}, {Tam}, {Van Dyk}, \&
  {Wheelock}}]{2mass}
{Skrutskie}, M.~F., {et~al.} 2006, \aj, 131, 1163

\bibitem[{{Solomon} {et~al.}(1987){Solomon}, {Rivolo}, {Barrett}, \&
  {Yahil}}]{solomon87}
{Solomon}, P.~M., {Rivolo}, A.~R., {Barrett}, J., \& {Yahil}, A. 1987, \apj,
  319, 730

\bibitem[{{Tsujimoto} {et~al.}(2007){Tsujimoto}, {Feigelson}, {Townsley},
  {Broos}, {Getman}, {Wang}, {Garmire}, {Baba}, {Nagayama}, {Tamura}, \&
  {Churchwell}}]{tsu07}
{Tsujimoto}, M., {et~al.} 2007, \apj, 665, 719

\bibitem[{{van den Bergh}(1978)}]{van78}
{van den Bergh}, S. 1978, \apjs, 38, 119

\bibitem[{{van Kerkwijk} {et~al.}(1995){van Kerkwijk}, {Kulkarni}, {Matthews},
  \& {Neugebauer}}]{vk95}
{van Kerkwijk}, M.~H., {Kulkarni}, S.~R., {Matthews}, K., \& {Neugebauer}, G.
  1995, \apjl, 444, L33

\bibitem[{{Vasisht} {et~al.}(1995){Vasisht}, {Frail}, \&
  {Kulkarni}}]{vasisht95}
{Vasisht}, G., {Frail}, D.~A., \& {Kulkarni}, S.~R. 1995, \apjl, 440, L65

\bibitem[{{Walsh} {et~al.}(1997){Walsh}, {Hyland}, {Robinson}, \&
  {Burton}}]{walsh97}
{Walsh}, A.~J., {Hyland}, A.~R., {Robinson}, G., \& {Burton}, M.~G. 1997,
  \mnras, 291, 261

\bibitem[{{Westerhout}(1958)}]{westerhout58}
{Westerhout}, G. 1958, \bain, 14, 215

\bibitem[{{Wilson} {et~al.}(1970){Wilson}, {Mezger}, {Gardner}, \&
  {Milne}}]{wilson70}
{Wilson}, T.~L., {Mezger}, P.~G., {Gardner}, F.~F., \& {Milne}, D.~K. 1970,
  \aap, 6, 364

\bibitem[{{Wink} {et~al.}(1982){Wink}, {Altenhoff}, \& {Mezger}}]{wink82}
{Wink}, J.~E., {Altenhoff}, W.~J., \& {Mezger}, P.~G. 1982, \aap, 108, 227

\bibitem[{{Wink} {et~al.}(1983){Wink}, {Wilson}, \& {Bieging}}]{wink83}
{Wink}, J.~E., {Wilson}, T.~L., \& {Bieging}, J.~H. 1983, \aap, 127, 211

\bibitem[{{Wood} \& {Churchwell}(1989)}]{wood89}
{Wood}, D.~O.~S., \& {Churchwell}, E. 1989, \apjs, 69, 831

\bibitem[{{Wyrowski} {et~al.}(2006){Wyrowski}, {Menten}, {Schilke},
  {Thorwirth}, {G{\"u}sten}, \& {Bergman}}]{wyr06}
{Wyrowski}, F., {Menten}, K.~M., {Schilke}, P., {Thorwirth}, S., {G{\"u}sten},
  R., \& {Bergman}, P. 2006, \aap, 454, L91

\bibitem[{{Xu} {et~al.}(2009){Xu}, {Reid}, {Menten}, {Brunthaler}, {Zheng}, \&
  {Moscadelli}}]{xu09}
{Xu}, Y., {Reid}, M.~J., {Menten}, K.~M., {Brunthaler}, A., {Zheng}, X.~W., \&
  {Moscadelli}, L. 2009, \apj, 693, 413

\end{thebibliography}

\clearpage

\begin{deluxetable}{ccccccccl}
  \tablewidth{0pt} \tabletypesize{\scriptsize} \tablecaption{WMAP
    Source List  \label{wmap}} \tablehead{ \colhead{} & \colhead{$l$} &
    \colhead{$b$} & \colhead{$smaj$} & \colhead{$smin$} &
    \colhead{$PA$} & \colhead{S$_{ff}$}
    & \colhead{Luminosity} & \colhead{} \\
    \colhead{Name} & \colhead{(deg)} & \colhead{(deg)} &
    \colhead{(deg)} & \colhead{(deg)} & \colhead{(deg)} &
    \colhead{(Jy)} & \colhead{Rank} & \colhead{Notes}}

  \startdata
  G10 & 10.4 & -0.3 & 0.61 & 0.44 & -21.2 & 86 & 12 & W31 \\

  G24 & 24.5 & 0 & 1.96 & 0.83 & -7.2 & 1377 & 10, 11 & W41, W42 \\

  G30 & 30.5 & 0 & 2.27 & 0.95 & 2.7 & 1585 & 3, 8 & W43 \\

  G34 & 34.7 & -0.2 & 0.92 & 0.57 & 6.1 & 285 & 6 & \nodata \\

  G37 & 37.6 & 0 & 0.8 & 0.71 & -59.9 & 244 & 13 & W47 \\

  G49 & 49.3 & -0.3 & 0.99 & 0.55 & -13.5 & 458 & 9 & W51 \\

  G283 & 283.9 & -0.6 & 1.37 & 0.78 & -32.1 & 848 & 15 & NGC 3199, RCW49, Partial GLIMPSE Coverage \\

  G291 & 291.2 & -0.7 & 1.04 & 0.75 & -31.1 & 688 & 14 & NGC 3603, NGC3576,  No GLIMPSE Coverage \\

  G298 & 298.4 & -0.4 & 0.91 & 0.73 & -37.3 & 313 & 5  & \nodata \\

  G311 & 311.6 & 0.1 & 1.72 & 0.93 & 6.5 & 766 & 16 & \nodata \\

  G327 & 327.5 & -0.2 & 1.55 & 0.83 & -10.7 & 943 & 17 &\nodata \\

  G332 & 332.9 & -0.3 & 1.56 & 0.93 & -3.3 & 1787 & 7 & \nodata \\

  G337 & 337.3 & -0.1 & 2.58 & 1.2 & -8.9 & 2239 & 2, 4 & \nodata \\

  \nodata & 359.9 & -0.1 & 1.2 & 0.51 & -3 & 1105 & 1 & Galactic Centre region, exluded from analysis \\

\enddata
\tablecomments{The luminosity rank indicates the ranking of the source
  with respect to the ionizing luminosity produced, with multiple
  ranks indicating sources that were divided in Paper I.  The
  free-free flux is measured in the W band at 90 GHz from the WMAP
  Free-Free Foreground Emission Map \citep{wmappaper}.}
\end{deluxetable}

\clearpage

\begin{deluxetable}{ccccccccccccc}
  \tablewidth{0pt} \tabletypesize{\footnotesize}
  \tablecaption{Star Forming Region Parameters \label{sfrlist}}
  \tablehead{ \colhead{} & \colhead{$l$} & \colhead{$b$} &
    \colhead{smaj} & \colhead{smin} & \colhead{$PA$} &
    \colhead{$v_{LSR}$}& \colhead{$\Delta v_{m}$} & \colhead{$\Delta
      v_{c}$} &
    \colhead{$D$} & \colhead{$<$R$>$} & \colhead{$t_{dyn}$} & \colhead{KDA} \\
    \colhead{\#} & \colhead{(deg)} &\colhead{(deg)}
    &\colhead{(arcmin)} &\colhead{(arcmin)} &\colhead{(deg)} &
    \colhead{(km s$^{-1}$)} &\colhead{(km s$^{-1}$)} &\colhead{(km
      s$^{-1}$)} &\colhead{(kpc)} & \colhead{(pc)} & \colhead{(Myr)} &
    \colhead{Ref.}}

  \startdata
  1 & 10.156 & -0.384 & 7.4 & 4.3 & 27 & 15 & 15 & 24 & 14.5 & 25 & 1.0 & 3 \\
  2 & 10.288 & -0.136 & 6.8 & 4.4 & 89 & 10 & \nodata & \nodata & 15.2 & 25 & \nodata & 3 \\
  3 & 10.450 & 0.021 & 3.3 & 2.4 & 54 & 69 & \nodata & \nodata & 6.1 & 5 & \nodata & 1 \\
  4 & 10.763 & -0.498 & 11.7 & 7.3 & 55 & -1 & \nodata & \nodata & 16.7 & 46 & \nodata & 1 \\
  \hline
  5 & 22.991 & -0.345 & 17.3 & 14.2 & -16 & 76 & \nodata & \nodata & 4.9, 10.8 & 22, 49 & \nodata & \nodata \\
  6 & 23.443 & -0.237 & 2.1 & 1.8 & 90 & 104 & \nodata & \nodata & 9.2 & 5 & \nodata & 1 \\
  7 & 23.846 & 0.152 & 8.9 & 4.9 & 8 & 95 & 17 & 27 & 5.8, 9.8 & 12, 20 & 0.4, 0.7 & \nodata \\
  8 & 24.050 & -0.321 & 26.1 & 14.5 & -9 & 85 & 17 & 24 & 10.3 & 61 & 2.6 & 3 \\
  9 & 24.133 & 0.438 & 2.9 & 2.3 & 0 & 98 & \nodata & \nodata & 5.9, 9.7 & 4, 7 & \nodata & \nodata \\
  10 & 24.911 & 0.134 & 26.7 & 21.8 & 0 & 100 & 15 & 21 & 6.1 & 43 & 2.1 & 3 \\
  11 & 25.329 & -0.275 & 7.5 & 5.0 & 90 & 63 & 4 & 8 & 4.1 & 7 & 0.9 & 7 \\
  12 & 25.992 & 0.119 & 18.4 & 11.4 & 0 & 106 & 4 & \nodata & 6.5, 8.8 & 28, 38 & 7.0, 9.5 & \nodata \\
  \hline
  13 & 28.827 & -0.230 & 1.4 & 1.1 & 22 & 88 & 5 & \nodata & 5.3 & 2 & 0.4 & 4 \\
  14 & 29.926 & -0.049 & 5.4 & 5.2 & 0 & 97 & \nodata & \nodata & 8.7 & 13 & \nodata & 1 \\
  15 & 30.456 & 0.443 & 3.0 & 2.6 & 0 & 58 & \nodata & \nodata & 3.6, 11 & 3, 9 & \nodata & \nodata \\
  16 & 30.540 & 0.022 & 1.0 & 0.8 & -48 & 44 & 5 & \nodata & 11.9 & 3 & 0.6 & 4 \\
  17 & 30.590 & -0.024 & 23.9 & 14.3 & 0 & 99 & 7 & 8 & 6.3 & 35 & 4.2 & 2 \\
  18 & 32.162 & 0.038 & 5.3 & 4.7 & 0 & 96 & \nodata & \nodata & 8.2 & 12 & \nodata & 2 \\
  \hline
  19 & 34.243 & 0.146 & 9.2 & 7.1 & -13 & 37 & \nodata & \nodata & 2.2 & 5 & \nodata & 2 \\
  20 & 35.038 & -0.490 & 3.8 & 2.4 & 76 & 51 & \nodata & \nodata & 3 & 3 & \nodata & 1 \\
  21 & 35.289 & -0.073 & 26.6 & 12.5 & 0 & 48 & \nodata & \nodata & 11 & 62 & \nodata & 2 \\
  \hline
  22 & 37.481 & -0.384 & 24.8 & 23.8 & 0 & 50 & 13 & 17 & 10.5 & 74 & 4.4 & 2 \\
  23 & 38.292 & -0.021 & 9.8 & 9.4 & 0 & 57 & \nodata & \nodata & 3.5, 9.9 & 10, 28 & \nodata & \nodata \\
  \hline
  24 & 49.083 & -0.306 & 8.9 & 4.7 & 0 & 68 & 4 & 6 & 5.6 & 11 & 2.0 & \nodata \\
  25 & 49.483 & -0.343 & 13.9 & 8.0 & 69 & 60 & 10 & 13 & 5.7 & 18 & 1.4 & \nodata \\
  \hline
  26 & 283.883 & -0.609 & 66.6 & 41.3 & -28 & 0 & 15 & 20 & 4 & 63 & 3.2 & 5 \\
  \hline
  27 & 290.873 & -0.742 & 36.6 & 25.8 & 90 & -22 & 6 & 10 & 3 & 27 & 2.8 & \nodata \\
  28 & 291.563 & -0.569 & 34.4 & 27.4 & -25 & 16 & 7 & 11 & 7.4 & 67 & 6.3 & \nodata \\
  \hline
  29 & 298.505 & -0.522 & 24.7 & 24.2 & 0 & 24 & 7 & 11 & 9.7 & 69 & 6.2 & \nodata \\
  \hline
  30 & 310.985 & 0.409 & 3.2 & 2.8 & 90 & -51 & \nodata & \nodata & 3.5 & 3 & \nodata & 6 \\
  31 & 311.513 & -0.027 & 27.5 & 25.0 & 90 & -55 & 10 & 13 & 7.4 & 56 & 4.3 & 6 \\
  32 & 311.650 & -0.528 & 9.0 & 5.4 & 90 & 34 & \nodata & \nodata & 13.6 & 29 & \nodata & \nodata \\
  \hline
  33 & 327.436 & -0.058 & 34.9 & 31.7 & 0 & -60 & 13 & 17 & 3.7 & 36 & 2.1 & 6 \\
  \hline
  34 & 332.809 & -0.132 & 48.6 & 30.2 & 0 & -52 & 6 & 7 & 3.4 & 39 & 5.9 & 6 \\
  35 & 333.158 & -0.076 & 2.5 & 1.9 & 90 & -91 & \nodata & \nodata & 5.5, 9.7 & 4, 6 & \nodata & \nodata \\
  \hline
  36 & 336.484 & -0.219 & 4.1 & 3.3 & -60 & -88 & 6 & 10 & 5.4, 10.2 & 6, 11 & 0.6, 1.2 & \nodata \\
  37 & 336.971 & -0.019 & 19.0 & 11.8 & 31 & -74 & 7 & 10 & 10.9 & 49 & 5.0 & 6 \\
  38 & 337.848 & -0.205 & 22.1 & 13.8 & -16 & -48 & 7 & 9 & 3.5 & 18 & 2.0 & 6 \\
  39 & 338.412 & 0.120 & 6.6 & 3.1 & 68 & -33 & 4 & 4 & 2.5, 13.3 & 4, 19 & 1.0, 4.8 & \nodata \\
  40 & 338.888 & 0.618 & 10.1 & 6.9 & 25 & -63 & \nodata & \nodata & 4.4 & 11 & \nodata & 6 \\

  \enddata
  
  \tablerefs{(1) \citet{churchwell90}; (2) \citet{kuchar94}; (3)
    \citet{sewilo04}; (4) \citet{palagi93}; (5) \citet{wilson70}; (6)
    \citet{caswell87}; (7) \citet{kolpak03}} \tablecomments{The KDA
    Ref indicates the source by which the kinematic distance ambiguity
    is resolved. In cases where no reference is given, the kinematic
    distance is either unique or we present the near and far
    distances.  }

\end{deluxetable}

\clearpage

\begin{deluxetable}{cccccccccc}
\tablewidth{0pt} \tabletypesize{\scriptsize} \tablecaption{Bubble-like Objects
Associated with the Star Forming Regions \label{bubblelist}}
\tablehead{\colhead{} &\colhead{$l$} &\colhead{$b$} &\colhead{smaj} &\colhead{smin} &\colhead{$PA$} 
&\colhead{Associated} &\colhead{$<$R$>$} &\colhead{Classification} &\colhead{GLIMPSE} \\
\colhead{\#} &\colhead{(deg)} &\colhead{(deg)} &\colhead{(arcmin)} &\colhead{(arcmin)} &\colhead{(deg)} 
&\colhead{SF Region} & \colhead{ (pc) } &\colhead{Flag} &\colhead{Bubble} }

\startdata
1 & 10.316 & -0.137 & 1.6 & 1.2 & 50 & 2 & 6.1 & C & \nodata \\
2 & 10.763 & -0.498 & 11.7 & 7.3 & 55 & 4 & 46 & B & N2 \\
3 & 23.703 & 0.166 & 0.3 & 0.3 & 66 & 7 & 0.5, 0.9 & C & \nodata \\
4 & 23.873 & -0.114 & 0.6 & 0.4 & -76 & 8 & 1.5 & C & \nodata \\
5 & 23.880 & -0.350 & 0.4 & 0.3 & 69 & 8 & 1.0 & C & \nodata \\
6 & 23.904 & 0.070 & 0.4 & 0.4 & 90 & 7 & 0.7, 1.2 & C & N32 \\
7 & 24.050 & -0.321 & 26.1 & 14.5 & -9 & 8 & 61 & C & \nodata \\
8 & 24.095 & 0.456 & 0.8 & 0.7 & -52 & 9 & 1.3, 2.1 & C & \nodata \\
9 & 24.133 & 0.438 & 2.9 & 2.3 & 0 & 9 & 4.5, 7.3 & B & \nodata \\
10 & 24.505 & 0.240 & 5.4 & 3.1 & -50 & 10 & 7.5 & B & N35 \\
11 & 24.840 & 0.106 & 4.6 & 2.1 & -64 & 10 & 5.9 & B & N36 \\
12 & 25.292 & 0.296 & 2.2 & 1.5 & 90 & 10 & 3.3 & B & N37 \\
13 & 25.323 & -0.293 & 5.4 & 5.4 & 0 & 11 & 6.4 & B & \nodata \\
14 & 25.992 & 0.119 & 18.4 & 11.4 & 0 & 12 & 28, 38 & B & \nodata \\
15 & 26.112 & -0.030 & 2.4 & 2.0 & 0 & 12 & 4.2, 5.6 & B & \nodata \\
16 & 28.827 & -0.230 & 1.4 & 1.1 & 22 & 13 & 1.9 & C & N49 \\
17 & 30.456 & 0.443 & 3.0 & 2.6 & 0 & 15 & 2.9, 9.0 & C & \nodata \\
18 & 30.612 & -0.194 & 11.0 & 5.8 & 0 & 17 & 15 & B & \nodata \\
19 & 30.736 & -0.022 & 3.0 & 2.1 & -36 & 17 & 4.7 & C & N52 \\
20 & 32.098 & 0.090 & 1.0 & 0.8 & -43 & 18 & 2.1 & C & N55 \\
21 & 32.159 & 0.039 & 4.6 & 4.0 & 0 & 18 & 10 & C & \nodata \\
22 & 34.156 & 0.145 & 3.8 & 2.6 & 11 & 19 & 2.0 & B & N61 \\
23 & 34.296 & 0.077 & 2.0 & 1.9 & 90 & 19 & 1.2 & B & \nodata \\
24 & 34.331 & 0.213 & 1.7 & 1.5 & -38 & 19 & 1.0 & B & N62 \\
25 & 35.002 & -0.130 & 5.1 & 3.2 & -68 & 21 & 13 & B & \nodata \\
26 & 35.032 & -0.482 & 1.6 & 1.4 & 54 & 20 & 1.3 & B & \nodata \\
27 & 35.259 & 0.119 & 0.5 & 0.4 & -28 & 21 & 1.5 & C & N66 \\
28 & 35.642 & -0.059 & 5.8 & 2.9 & -75 & 21 & 14 & C & N68 \\
29 & 37.481 & -0.384 & 24.8 & 23.8 & 0 & 22 & 74 & B & \nodata \\
30 & 38.293 & -0.015 & 8.1 & 7.4 & 0 & 23 & 7.9, 22 & B & N71 \\
31 & 38.352 & -0.133 & 1.0 & 0.9 & 0 & 23 & 1.0, 2.8 & B & N72 \\
32 & 298.218 & -0.323 & 2.4 & 1.7 & 90 & 29 & 5.8 & C & S181 \\
33 & 298.527 & -0.562 & 18.5 & 14.7 & -51 & 29 & 47 & C & \nodata \\
34 & 310.985 & 0.409 & 3.2 & 2.8 & 90 & 30 & 3.1 & C & S137 \\
35 & 311.487 & 0.401 & 2.5 & 1.7 & 12 & 31 & 4.6 & B & S133 \\
36 & 311.495 & -0.013 & 24.2 & 17.7 & -72 & 31 & 45 & B & \nodata \\
37 & 311.916 & 0.222 & 2.2 & 1.4 & -73 & 31 & 3.9 & C & \nodata \\
38 & 326.882 & -0.389 & 13.1 & 8.0 & -66 & 33 & 11 & B & \nodata \\
39 & 327.532 & -0.605 & 14.9 & 14.1 & 0 & 33 & 16 & B & \nodata \\
40 & 333.158 & -0.076 & 2.5 & 1.9 & 90 & 35 & 3.6, 6.3 & B & \nodata \\
41 & 336.484 & -0.219 & 4.1 & 3.3 & -60 & 36 & 5.8, 11 & C & \nodata \\
42 & 336.778 & 0.089 & 2.9 & 2.4 & -20 & 37 & 8.4 & B & \nodata \\
43 & 336.853 & -0.016 & 1.6 & 1.2 & 82 & 37 & 4.5 & B & \nodata \\
44 & 336.887 & 0.059 & 2.7 & 1.6 & 0 & 37 & 6.8 & B & \nodata \\
45 & 336.921 & -0.195 & 2.2 & 1.9 & -70 & 37 & 6.4 & C & \nodata \\
46 & 337.153 & -0.178 & 1.8 & 1.6 & 68 & 37 & 5.4 & B & \nodata \\
47 & 337.676 & -0.335 & 2.1 & 1.8 & 35 & 38 & 2.0 & C & S37 \\
48 & 338.891 & 0.596 & 3.7 & 3.0 & -55 & 40 & 4.3 & B & S29 \\
\enddata

\tablecomments{The classification flag refers to the morphology 
of the bubble: 'C' indicates that the bubble is a closed bubble 
and 'B' refers to a broken bubble.\\
The GLIMPSE bubble column indicates the identification of this
bubble in \citet{churchwell06} with the catalogue number. }

\end{deluxetable}

\clearpage

\begin{deluxetable}{cccccc}
  \tablewidth{0pt} \tabletypesize{\scriptsize}
  \tablecaption{\ion{H}{2} Region Velocities\label{hiilist}}
  \tablehead{\colhead{Associated Star} & \colhead{} & \colhead{$l$}
    &\colhead{$b$} &\colhead{$v_{\textrm{LSR}}$}
    &\colhead{Reference} \\
    \colhead{Forming Region} & \colhead{Name} & \colhead{(deg)}
    &\colhead{(deg)} &\colhead{(km s$^{-1}$)} &\colhead{Number} }

  \startdata
  1 & [L89b] 10.073-00.412 &  10.0732 & -0.4119 & 13.6 & 1 \\
  \nodata & GRS 010.20 -00.30  & 10.1467 &  -0.3377 & 13 & 2  \\
  \nodata & [WMG70] 010.2-00.3  & 10.149 &  -0.343 & 14.2 & 3  \\
  \nodata & PMN J1809-2019  & 10.1589 &  -0.3489 & 14.2 & 1  \\
  \nodata & GAL 010.2-00.3  & 10.1615 &  -0.3546 & 12.8 & 4  \\
  \nodata & [L89b] 10.190-00.426  & 10.1898 &  -0.426 & 36.2 & 1  \\
  \nodata & [L89b] 10.190-00.426  & 10.1898 &  -0.426 & 5.4 & 1  \\
  2 & [WC89] 010.30-0.15B  & 10.303 &  -0.1462 & 7.7 & 5  \\
  \nodata & GAL 010.32-00.16  & 10.3148 &  -0.15 & 13.1 & 4  \\
  3 & GAL 010.46+00.02  & 10.458 &  0.0239 & 70.1 & 1  \\
  \nodata & GAL 010.46+00.03  & 10.4625 &  0.034 & 68.9 & 4  \\
  \nodata & GAL 010.47+00.03    & 10.4722 &  0.0264 & 68 & 6  \\
  4 & GAL 010.6-00.4  & 10.6205 &  -0.3872 & 0.2 & 7  \\
  \nodata & [SG70] 010.6-0.4  & 10.6258 &  -0.383 & 0.4 & 8  \\
  \nodata & [KC97c] G010.7-00.5  & 10.6639 &  -0.4666 & -2.4 & 1  \\
  \enddata

  \tablerefs{(1) \citet{lockman89}; (2) \citet{anan85}; (3)
    \citet{wilson70}; (4) \citet{bronfman96}; (5) \citet{kim01}; (6)
    \citet{wood89}; (7) \citet{wink82}; (8) \citet{lockman96}; (9)
    \citet{quireza06}; (10) \citet{kolpak03}; (11) \citet{caswell72};
    (12) \citet{wink83}; (13) \citet{pestalozzi05}; (14)
    \citet{kuchar94}; (15) \citet{walsh97}; (16) \citet{rr02}; (17)
    \citet{caswell87}; (18) \citet{georgelin70}; (19)
    \citet{russeil05}; (10) \citet{ikeda01}; (21) \citet{gardner78};
    (22) \citet{caswell75}; (23) \citet{downes80}}

  \tablecomments{Table \ref{hiilist} is published in its entirety in
    the electronic edition of the Astrophysical Journal. A portion is
    shown here for guidance regarding its form and content. }

\end{deluxetable}

\clearpage

\begin{deluxetable}{cccc}
  \tablewidth{0pt} \tabletypesize{\scriptsize} \tablecaption{Dynamical
    Properties of the Star Forming Regions \label{dynprop}}
  \tablehead{\colhead{SFR} & \colhead{$\log$ M$_{sh}$} & \colhead{$\log$ E$_{k}$}  & \colhead{$\log$ L$_{mech}$} \\
    \colhead{\#} & \colhead{($\textrm{M}_{\sun}$)} & \colhead{($
      \textrm{erg}$)} & \colhead{($ \textrm{erg s}^{-1} $)} }

\startdata
1  & 4.9 &  50.7  & 37.8 \\
2  & 4.9 &  50.1  & 36.9 \\
3  & 2.8 &  48.0  & 36.2 \\
4  & 5.7 &  50.9  & 37.2 \\
5  & 5.4 &  50.5  & 37.1 \\
6  & 2.9 &  48.1  & 36.2 \\
7  & 4.3 &  50.2  & 37.7 \\
8  & 6.1 &  51.8  & 38.1 \\
9  & 2.9 &  48.1  & 36.2 \\
10 & 5.6 &  51.3  & 37.8 \\
11 & 3.3 &  48.1  & 35.8 \\
12 & 5.3 &  49.5  & 35.6 \\
13 & 1.6 &  46.0  & 34.6 \\
14 & 4.1 &  49.3  & 36.6 \\
15 & 3.1 &  48.2  & 36.3 \\
16 & 2.2 &  46.6  & 34.8 \\
17 & 5.4 &  50.2  & 36.6 \\
18 & 4.0 &  49.1  & 36.6 \\
19 & 2.9 &  48.0  & 36.2 \\
20 & 2.0 &  47.2  & 35.9 \\
21 & 6.1 &  51.3  & 37.3 \\
22 & 6.3 &  51.8  & 37.8 \\
23 & 4.6 &  49.7  & 36.8 \\
24 & 3.9 &  48.3  & 35.5 \\
25 & 4.5 &  49.7  & 36.8 \\
26 & 6.1 &  51.7  & 37.9 \\
27 & 5.0 &  50.0  & 36.6 \\
28 & 6.2 &  51.2  & 37.1 \\
29 & 6.2 &  51.3  & 37.2 \\
30 & 2.2 &  47.3  & 36.0 \\
31 & 6.0 &  51.2  & 37.3 \\
32 & 5.1 &  50.3  & 37.0 \\
33 & 5.4 &  50.8  & 37.5 \\
34 & 5.5 &  50.1  & 36.3 \\
35 & 2.8 &  48.0  & 36.2 \\
36 & 3.5 &  48.5  & 36.1 \\
37 & 5.8 &  50.8  & 36.9 \\
38 & 4.5 &  49.4  & 36.4 \\
39 & 3.9 &  48.1  & 35.1 \\
40 & 3.8 &  49.0  & 36.5 \\
\enddata

\end{deluxetable}

\clearpage

\clearpage

\clearpage

\end{document}